\pdfoutput=1
\documentclass[12pt]{article}
\usepackage{amsfonts,palatino,amsthm}
\usepackage{amsmath}
\usepackage{amssymb}
\usepackage{epsfig}
\usepackage{epstopdf}
\usepackage{textcomp}
\usepackage{cancel}
\usepackage{wrapfig}
\usepackage{bbm}
\def\eali\end{align*}
\def\be{\begin{equation}}
\def\ee{\end{equation}}
\def\bea{\begin{eqnarray}}
\def\eea{\end{eqnarray}}
\def\ba{\begin{array}}
\def\ea{\end{array}}
\def\bali{\begin{align*}}

\def\bite{\begin{itemize}}
\def\chew{\item}
\def\eat{\end{itemize}}
\def\begfig{\begin{figure}[h!]}
\def\endfig{\end{figure}}
\def\capt#1{\caption{{{#1}}}}

\def\ket#1{\mid #1 {\cal{i}}}

\def\N{\ensuremath{\mathbbm{N}}}

\def\R{\ensuremath{\mathbbm{R}}}

\def\fg#1{Figure \ref{#1}}
\def\bd{\bf \begin{definition}: \it}
\def\ed{\end{definition} \rm}
\def\blem{\bf \begin{lemma}: \it}
\def\elem{\end{lemma} \rm}
\def\bthe{\bf \begin{theorem}: \it}
\def\ethe{\end{theorem} \rm}
\def\bcor{\bf \begin{corollary}: \it}
\def\ecor{\end{corollary} \rm}
\def\bpro{\bf \begin{proof}: \rm}
\def\epro{\end{proof} \rm}

\def\su2{\ensuremath{\mathfrak{su(2)}}}

\def\+{\ensuremath{\ket{+}}}
\def\-{\ensuremath{\ket{-}}}

\def\ad{\ensuremath{\dagg{\mathbf{a}}}}

\def\bb{2$\rightarrow$2}
\def\ac{1$\rightarrow$3}
\def\ca{3$\rightarrow$1}

\def\ad{1$\rightarrow$4}

\begin{document}
\title{Conserved Topological Defects in Non-Embedded Graphs in Quantum Gravity}
\author{Fotini Markopoulou and Isabeau Pr\'emont-Schwarz \\ \\ Perimeter Institute for Theoretical Physics \\ Waterloo, Ontario N2L 2Y5, Canada \\ \\ and \\ \\ University of Waterloo \\ Waterloo, Ontario N2L 3G1, Canada}
\date{May 2, 2008}

\maketitle
\begin{abstract}
We follow up on previous work which found
that commonly used graph evolution moves lead to conserved quantities that can be expressed in terms of the braiding of the graph in its embedding space.
We study non-embedded graphs under three distinct sets of dynamical rules and find non-trivial conserved quantities that can be expressed in terms of topological defects in the dual geometry.
 For graphs dual to 2-dimensional simplicial complexes we identify all the conserved quantities of the evolution. We also indicate expected results for graphs dual to 3-dimensional simplicial complexes.
\end{abstract}
\newpage
\tableofcontents
\newpage
%%%%%%%%%%%%%%%%%%%%%%%%%%%%%%%%%%%%%%%%%%%%%%%%%%%%%%%%%%%%%%%%%%%%%%%%%%%%%%%%
%--I N T R O----I N T R O----I N T R O----I N T R O----I N T R O----I N T R O--%
%%%%%%%%%%%%%%%%%%%%%%%%%%%%%%%%%%%%%%%%%%%%%%%%%%%%%%%%%%%%%%%%%%%%%%%%%%%%%%%%
\section{Introduction}

Background independence in a quantum theory of gravity is the requirement that the physical content of the theory does not depend on a given geometry.  This is postulated to be a desirable feature, mainly justified by analogy with the classical theory of gravity, general relativity, which is a theory of dynamical spacetime geometry.
Background independent approaches to quantum gravity include Loop Quantum Gravity \cite{Thie02,AshLew04,RovBook,Pul,Kie,Nic}, Spin Foams \cite{Ori03,Per04} and Causal Sets \cite{Sor03,Dow05}.   For a review of background independence in quantum gravity, see \cite{RovBook,ButIsh99,Smo05,Mar07b}.

Probably the biggest open problem facing background independent approaches to quantum gravity is what we shall call the low energy problem, namely, showing that a given candidate microscopic theory at low energy reproduces the known low energy physics, general relativity coupled to quantum fields.  The problem is not whether the microscopic theory has the correct limit or not, it is in taking the limit.  While background independence has many appealing features, it makes difficult the use of standard tools for taking such a limit (see, for example, \cite{Mar07a}).

Some progress has been made in this direction in Loop Quantum Gravity using coherent states \cite{Thi00,Thi02,SahThi02,AshBomCor05,BahThi07} and searching for the graviton propagator \cite{BlaModRovSpe}.  It is also possible to obtain substantial results in 2+1 LQG and spin foams (see, e.g., {\mbox{\cite{3dgravityLL, 3dgravityOT, 3dgravityN}}), however, this is largely because of the special topological nature of the theory in 2+1 dimensions and has not yet been extended to higher dimensions.

An additional direction was recently proposed in \cite{KriMar05,Mar07a} who suggested  that a first step towards the low energy limit of a background independent quantum theory of gravity is to search for conserved quantities in the microscopic theory.  The basic idea is that the properties and symmetries of such conserved quantities will be present in the low energy theory and can be used to characterize it.

In \cite{DreMarSmo06,KriMar05,Mar07a}, we proposed using methods from quantum information theory, such as noiseless subsystems \cite{ZanRas97,KemBacLidWha01,HolKriLaf03,KriLafPou04} to identify such conserved quantities.  The method is promising because it can readily be applied to a large class of background independent theories.  Using this method, in \cite{DreMarSmo06,BilMarSmo07} we showed that such quantities do exist under commonly considered dynamics (evolution of networks/spin networks under local network moves).   In \cite{BilMarSmo07}, the conserved topological quantities were speculated to be quantum numbers of matter degrees of freedom, by correspondence to the preon model of \cite{Bil05}.

The idea that matter can be encoded in topological defects of the geometry is an old one (see, for example, \cite{Whe57,FinMis59,BriHar64,PerCoo99}).
 Later work \cite{SmoWan07,HacWan08,BilHacKauSmo08,HeWan08a,HeWan08b} has studied in more detail the form of the conserved braiding and the propagation of braids in the graph states.  Another possibility is that the conserved quantities, exhibiting different local topologies can, in conjuction with some labeling of the graph, correspond to topological geons, which themselves have the freedom of constituting matter degrees of freedom since they can exibit both fermionic and bosonic spin statistics \mbox{\cite{Dowker:1996ei}}.

At the kinematical level, there are two broad classes of network states that have been considered in the quantum gravity literature:  networks embedded in a manifold, and non-embedded, or abstract, networks.  In the latter case, only the combinatorial information of the connectivity of the network is relevant for the physics.  While the former is relevant in Loop Quantum Gravity, the latter also regularly appear in related literature.  Abstract network states are attractive for a number of reasons.  Conceptually, since the network states are more fundamental than the embedding space, the latter seems like an unnatural remnant of the classical theory that has been quantized.  In fact, when spatial diffeomorphisms are taken into account in the construction of spin network states, much of the information about the embedding becomes irrelevant (see \cite{RovBook}).  Further reasons to consider abstract graphs in Loop Quantum Gravity are given in the new approach of Algebraic Loop Quantum Gravity \cite{GieThi06,GieThi07}.  Abstract networks have also been used in both early work on spin foam models (e.g., \mbox{\cite{Mar97,MarSmo97}}) and the most recent developments in spin foam models (\mbox{\cite{Engle:2007uq, Freidel:2007py, Oriti:2007vf}}). Finally, group field theory (see \mbox{\cite{Fre05, Oriti:2006se}}), one of the major approaches to quantum gravity,  uses exclusively abstract graphs dual to a simplicial complex.

In this article, we return to the question of the choice between abstract and embedded networks from the new viewpoint of the search for conserved quantities under the graph evolution moves.  The recent work that has motivated the present article used a state space of embedded graphs.  They found that what is conserved is the braiding of the edges of the graphs.  Since there is an infinite number of braids for a finite set of graph edges, there is an infinite set of conserved quantities.  It is not clear at this stage what the physical information encoded by these braids is, however, if we believe the scenario of \cite{Bil05,BilMarSmo07}, this implies  an infinite number of particle generations.  As this may be an artifact of the embedding space and, as we discussed above, the embedding information may not be relevant at the fundamental level, we here wish to study the conserved quantities in the case of non-embedded graphs.

One may worry that since braiding of an abstract graph can be undone, the result will be trivial.
In fact, we find not only that there are non-trivial topological defects that are conserved but that they can be nicely characterized, at least in the 2-dimensional case, and there are advantages over the embedded case:  there is a finite number of conserved quantities for a finite graph and there are no ambiguities in the conserved quantities resulting from the embedding topology.
We study three classes of dynamics, ordered
by an increasing number of dynamical rules.  We find a corresponding hierarchy of conserved quantities, ordered by the inclusion relationship, with more conserved topological defects when there are more dynamical constraints.  We note that the rule with the most conserved quantities has an infinite number of these but they can all be constructed by the same finite building block.   This rule also has some conserved quantities that are environment dependent.

In short, we find that commonly used graph evolution moves lead to conserved quantities that can be expressed in terms of topological defects in the dual geometry and in two dimensions we give a characterization of these for three sets of evolution rules.  These have not been seen before and, since they contain a substantial part of the graph information, we expect that they are of physical significance.  We hope that their physical meaning can be understood in future work,

The outline of this article is as follows.  In Section 2, we give the basic definitions of the graphs we consider and the construction of the dual geometry.  Section 3 gives the dynamics of the three different models we study and makes explicit the hierarchy between them and the ordering, with respect to the inclusion relation, of their conserved quantities.  In Section 4, we find the conserved quantities for each of the models in the non-embedded case and we comment on the embedded analogue.  In Section 5, we make some observations relevant for the interpretation of the conserved quantities as matter particles.  We summarize our conclusions in Section 6.  Some relevant technical details are presented in Appendices A-C.

%%%%%%%%%%%%%%%%%%%%%%%%%%%%%%%%%%%%%%%%%%%%%%%%%%%%%%%%%%%%%%%%%%%%%%%%%%%%%%%%%%%%%%%%%%
%--F O R M A L I S M ----F O R M A L I S M ----F O R M A L I S M ----F O R M A L I S M --%
%%%%%%%%%%%%%%%%%%%%%%%%%%%%%%%%%%%%%%%%%%%%%%%%%%%%%%%%%%%%%%%%%%%%%%%%%%%%%%%%%%%%%%%%%%
\section{Line Graphs}
\label{SectionLine}

We will be considering three types of mathematical objects which, in
the non-embedded case are all equivalent: line graphs, framed (or ribbon)
graphs and simplicial complexes.

An  {\em $n$-line graph} is composed of nodes and legs (edges) that connect nodes. A {\em leg} of an $n$-line graph contains n {\em lines}. At each node, $n+1$ legs meet such that each line in one leg is connected to a line in a different leg (\fg{dualex}).

 A {\em simplicial complex} is a topological space constructed by gluing together simplices. A $k$-simplex is the $k$-dimensional analog of a triangle, so a 0-simplex is a point, a 1-simplex is a segment, a 2-simplex is a triangle, a 3-simplex is a tetrahedron, etc. More generally, a $k$-simplex can be visualised as the convex hull of $n+1$ independent points. A face of a $k$-simplex $K$ is an $l$-simplex at the boundary of $K$ that is formed by taking the convex hull of a subset of order $l+1$ of the $k+1$ points whose convex hull is $K$ and where $l$ can take any value between 0 and $n$. A simplicial complex, C, satisfies two axioms: (1) if a $k$-simplex $K$ is in C, then all the faces of $K$ are also in C and (2)  the intersection of two simplices $K_1\ $ and $K_2 \ $ is a face of both $K_1\ $ and $K_2$.  An $n$-dimensional simplicial complex is a topological space constructed by gluing together $k$-simplices where $k$ is less than or equal to $n$. A {\em simplicial manifold}, is a simplicial complex which is also a piece-wise linear manifold.

For $n = 2$, there is an exact bijection between $n$-line graphs and $n$-dimensional simplicial manifolds. For n$>$2, there is an injection of $n$-dimensional simplicial manifolds into the set of $n$-line graphs but a generic $n$-line graph corresponds only to a pseudo-manifold as they contain topological defects of a different type as the ones which we will examine in this paper: they contain a subset of points (of measure zero) that have locally non-trivial topology (i.e. given a neighbourhood of the point, one cannot always find an n-ball containing the point but also contained in the neighbourhood)\mbox{\cite{DePietri:2000ii}}. It has been suggested that this other type of topological defect (called ``conical defect" but not to be confused with geometrical ``conical defects") may also correspond to matter degrees of freedom \mbox{\cite{Crane:2001kf}} . This correspondence between $n$-line graphs and pseudo-simplicial manifolds (or simplicial manifolds when n=2) defines a duality between these two mathematical structures.

The duality between the line graphs and the (pseudo) - simplicial manifolds
is as follows. Suppose we have a $n$-line graph, then
each node can be mapped to an $n$-dimensional
simplex. Each leg emerging from the node can be mapped to an $n-1$
dimensional face of the $n$-dimensional volume simplex and every
line in a leg represents an $(n-2)$-dimensional simplex bordering
the $n-1$ simplex represented by the leg. For example, in \fg{dualex},
which illustrates the $n=3$ case, a tetrahedron and its dual graph are
drawn, and to make the duality more explicit, the lines and their
respective dual edges are identically coloured. We draw the dual
graph by putting each node at the centre of its dual $n$-dimensional
simplex and by making each leg cross the $(n-1)$-dimensional simplex
dual to it. This will result in having the lines encircle the
$(n-2)$-dimensional simplex it is dual to.
For $n\in\{2,3\}$ the graph so constructed provides all the
necessary information of how the $n$-dimensional simplices are
glued together in the dual picture.

\begin{figure}
\begin{center}
\scalebox{1.3}{
\includegraphics{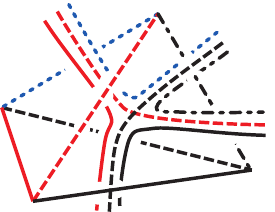}}
\end{center}
\caption{In the case $n=3$, the basic volume element is a
  tetrahedron. In the above tetrahedron and its dual graph, each graph leg is dual to a face of the
  tetrahedron and each line is dual to the edge of tetrahedron of
  the same colour.}
\label{dualex}
\endfig

Finally, a framed graph is an $n+1$ valent graph such that all the edges are
``thickened" so as to take the shape of an $(n-1)$-dimensional simplex
tensor product $\R$ (a triangular prism for $n=3$, as shown in \fg{tube}, and a ribbon for $n=2$). The
vertices are ``thickened" into $n$-simplices such that each of their
$(n-1)$-simplices join with a ``thickened edge".

%%%%%%%%%%%%%%%%%%%%%%%%%%%%%%%%%%%%%%%%%%%%%%%%%%%%%%%%%%%%%%%2D CASE%%%%%%%%%%%%%
\subsection{2d Case}
If we restrict ourselves to the case $n=2$, then the nodes are dual to
triangles, the legs dual to the edges and the lines dual to
points and the line graphs are exactly dual to simplicial manifolds. An example of a triangulation and its dual graph can be seen in
\fg{dual2d} where a ``flat'' layout is presented.  Each node sits at
the center of its dual triangle, each leg crosses its dual edge and
each line circles the point it is dual to.

\begin{figure}
\begin{center}
\begin{equation}
a)
\scalebox{.5}{
\includegraphics{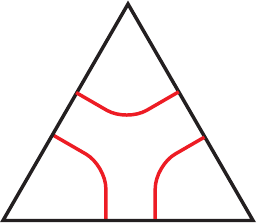}}
\qquad
b)
\scalebox{.5}{
\includegraphics{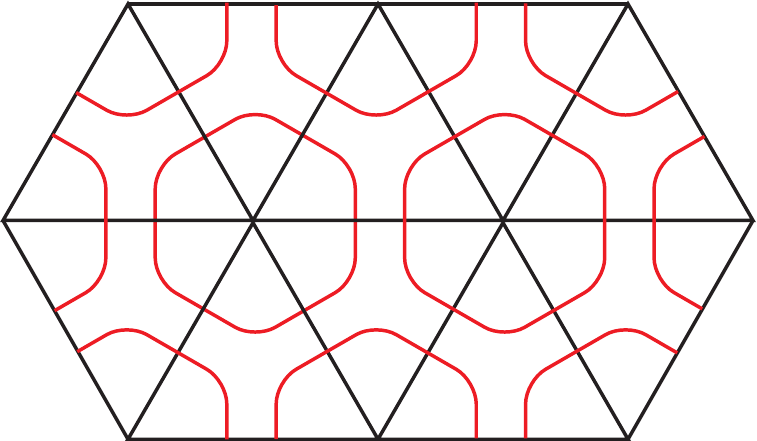}}
\nonumber
\end{equation}
\end{center}
\capt{In 2d, the basic volume element is a
  triangle. Here is a ``flat'' layout of triangles with its dual graph. The nodes are
  dual to the triangles; each leg is dual to an edge of a triangle
  and is composed of two lines themselves dual to the point they encircle. }
\label{dual2d}
\endfig

An interesting thing to note is that, if we consider the lines of the graph to
  be edges of a ribbon, each triangle is dual to a three-valent
  ribbon-vertex and the whole simplicial manifold corresponds to a trivalent
  \emph{framed} graph (such a graph is what is used by \cite{MarSmo97}
  though in their case it is embedded in a 3-manifold and has a
  different interpretation). If each line is shrunk to a point, then each
  trivalent ribbon vertex becomes homeomorphic to the triangle it
  is dual to and, furthermore, they are glued together identically like
  their dual triangles are. Hence, the ribbon graph is a
  2d surface full of holes whose boundaries, the lines, correspond
  to the vertices of the dual triangles. If these holes are removed by
  contracting their borders (the lines) to a point, then the 2-surface becomes homeomorphic to
  the dual triangulated surface.

\subsection{3d Case}
In the case $n=3$, the nodes are dual to tetrahedra, the legs dual
to their faces and the lines dual to their edges. As usual, the
duality can be obtained by placing the nodes at the center of
their dual tetrahedra and the legs crossing their dual faces. Each
line loops around its dual edge.

The framed graph in the $n=3$ case is obtained by considering the three lines of a leg to form a
  tube as shown in \fg{tube}.
  One then obtains 4-valent tube graphs. This formalism is used
  by \cite{SmoWan07}. We have an analogous result for the tube graph as we did for the ribbon graph, namely that, if the boundaries of the tubes (2d-surfaces)
  are shrunk down in the direction of the length of the tube to a
  single line or edge, then it becomes homeomorphic to the dual
  triangulation.

\begin{figure}
\begin{center}
\scalebox{.7}{
\includegraphics{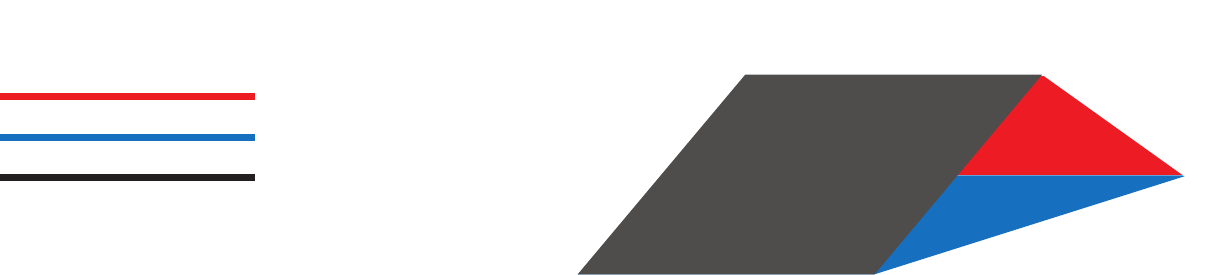}}
\end{center}
\capt{Instead of being represented by three lines, each leg can be
represented by a tube.} \label{tube}
\endfig

%%%%%%%%%%%%%%%%%%%%%%%%%%%%%%%%%%%%%%%%%%%%%%%%%%%%%%%%%%%%%%%%%%%%%%%%%%%%%%%%%%%%%%%%%%%%%%%%%%%%%%%%%%%
%%%%%%%%%%%% D Y N A M I C S    D Y N A M I C S    D Y N A M I C S    D Y N A M I C S    D Y N A M I C S %%
%%%%%%%%%%%%%%%%%%%%%%%%%%%%%%%%%%%%%%%%%%%%%%%%%%%%%%%%%%%%%%%%%%%%%%%%%%%%%%%%%%%%%%%%%%%%%%%%%%%%%%%%%%%
\section{Dynamics on Line Graphs}
\label{SectionDynamics}

The dynamics for the $n$-simplicial complex can be given by the Pachner moves, for example as suggested in \cite{ReiRov97,Mar97,MarSmo97,BaezSF}. The dynamics for simplicial complexes induces dynamics for line graphs via the duality between the two. The
Pachner moves for an $n$-simplicial complex consist of identifying a finite subset, $P$, of $n$-simplicies of the $n$-simplicial complex, C,  with a part $B_p\ $ of the boundary, $B$, of an $(n+1)$-simplex, $S$, and then replacing $P$ in C by ($B-B_p$). So for example, in the 2d case, the
Pachner moves consist of taking $k\in \{1,2,3\}$ triangles which are
in the same configuration as $k$ triangles of the boundary of a
tetrahedron and replacing them by the remaining $(4 - k)$ triangles of the
tetrahedron. The  four possible moves in 2d are shown in \fg{2dyn},
and the 3d moves in \fg{3dyn}.

\begin{figure}
\begin{center}
\begin{equation}
a)
\scalebox{.7}{
\includegraphics{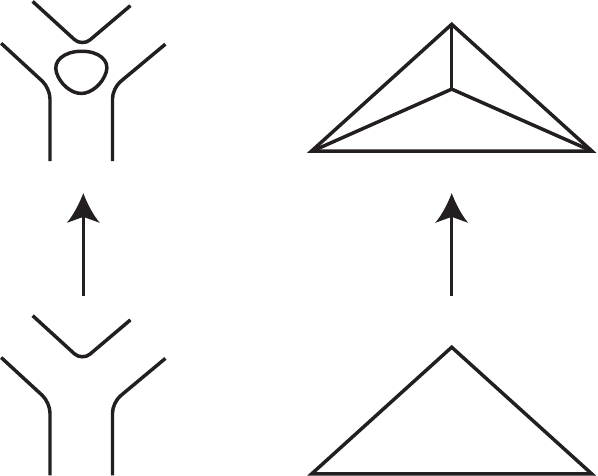}}
\qquad
b)
\scalebox{.7}{
\includegraphics{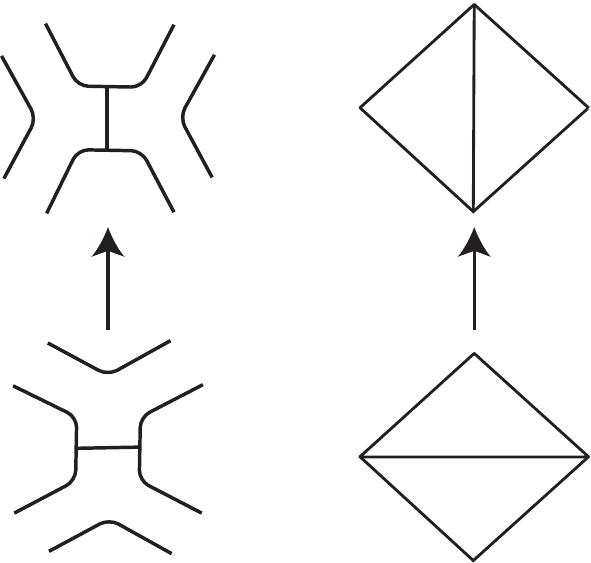}}
c)
\scalebox{.7}{
\includegraphics{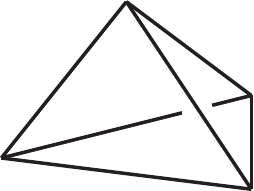}}
\nonumber
\end{equation}
\end{center}
\capt{a) The $1 \rightarrow 3$ graph move and the equivalent $1\rightarrow 3$
Pachner move. b) The $2\rightarrow 2$ graph move and the equivalent Pachner move. c) In both cases, the
 initial and final state form a 3-simplex
(tetrahedron).}
\label{2dyn}
\endfig

\begin{figure}
\begin{center}
\begin{equation}
\begin{array}{lclc}
a)&
\scalebox{.7}{
\includegraphics{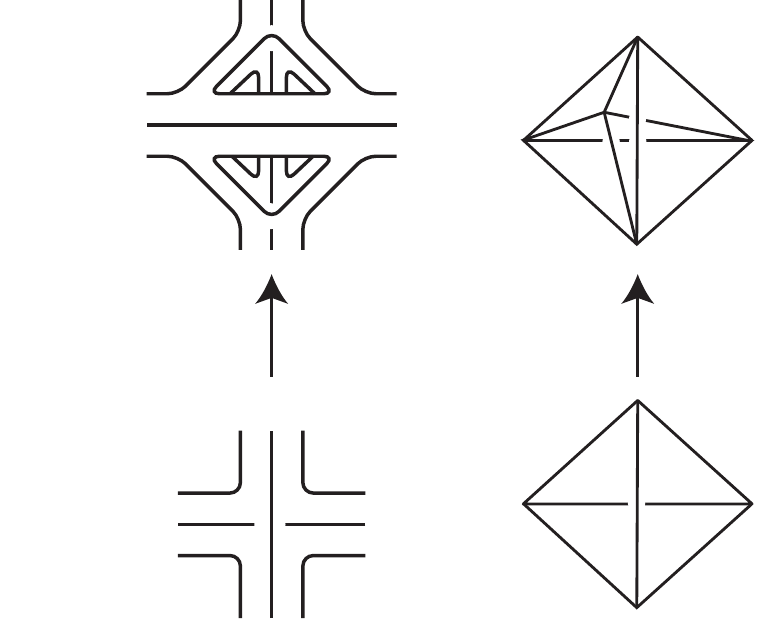}}&
b)&
\scalebox{.7}{
\includegraphics{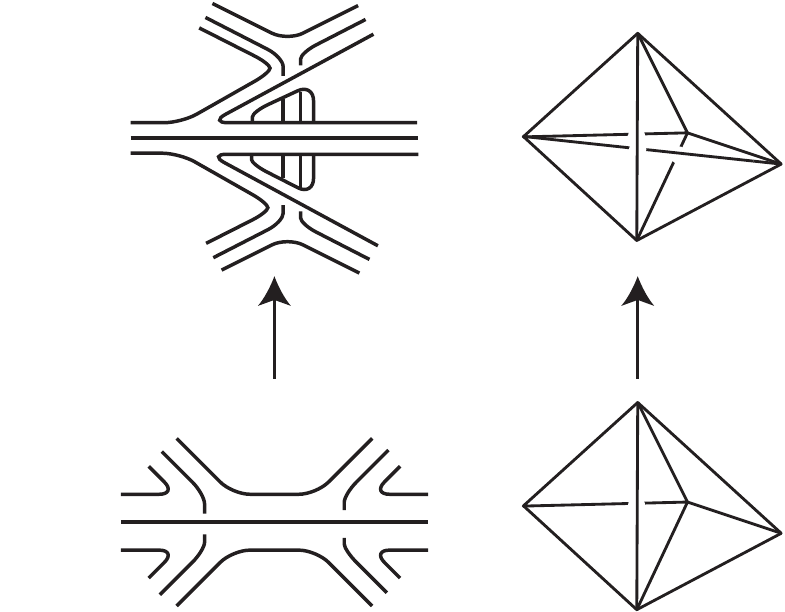}}\\
&&&\\
c)&
\scalebox{.7}{
\includegraphics{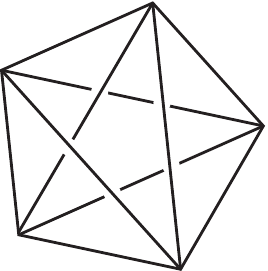}}&
&
\end{array}
\nonumber
\end{equation}
\end{center}
\capt{a) The $1 \rightarrow 4$ graph move and the equivalent $1\rightarrow 4$
Pachner move. b) The $2\rightarrow 3$ graph move and the equivalent Pachner move. c) In both cases, the
 initial and final state form a 4-simplex.}
\label{3dyn}
\endfig

Intuitively, the reason for using Pachner moves is that, by using them, the
time evolution should build up $(n+1)$-simplices which are
connected together to form an $(n+1)$-simplicial complex. One would thus expect these moves
to create a $(n+1)$-dimensional manifold from $n$-dimensional slices, though this not necessarily the case in practice, a counter-example is shown in \cite{fractal},  where fractal structures occur.

A possible incovenience of such dynamics is that the Pachner
moves leave the topology of the simplicial complex invariant. In fact, Pachner proved in 1987
that two closed $n$-simplicial manifolds are homomorphic
if and only if one can be transformed to the other by a sequence of
Pachner moves\cite{pach}. The use of Pachner moves for the dynamics therefore
restricts us to evolutions that do not change the topology. But general relativity, which we expect to be reproduced in an appropriate limit of the theory, does not have this restriction: a 3-geometry can evolve into a topologically distinct manifold.

\begin{figure}
\begin{center}
\scalebox{.9}{
\includegraphics{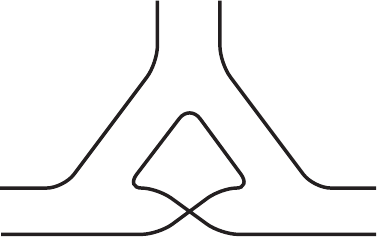}}
\end{center}
\capt{Even though this subgraph represents three triangular
simplices glued in such a manner as to form a larger triangle
bordered by three distinct edges connecting three distinct points
it cannot be contracted to a single simplex triangle through some
sort of  3$\rightarrow$1 Pachner move because it contains a
cross-cap and a single simplex triangle with three distinct edges
cannot contain a cross-cap.} \label{2duncontra}
\endfig

It is worth mentioning that,
viewed in the graph representation, Pachner moves have non-trivial
restrictions. At least in 2 and 3 dimensions, all the expanding moves (i.e., the Pachner moves that increase the total number of
triangles or tetrahedra) can be performed
without restrictions. The contracting moves are less trivial
though.
For example, in the 2d case, the diagram in \fg{2duncontra} cannot
be contracted with a 3$\rightarrow$1 move because it lacks a
closed line. Less trivially, the diagram in \fg{3duncontra} cannot
be contracted by a 3$\rightarrow$2 even though it looks like it
could be because it does not correspond to the required
configuration: it is not homeomorphic to three tetrahedra forming
part of the boundary of a 4-simplex.

\begin{figure}
\begin{center}
\scalebox{1}{
\includegraphics{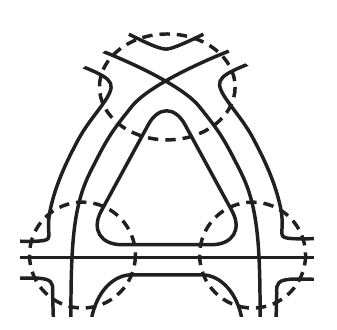}}
\end{center}
\capt{Even though this subgraph looks as if it may be
contractible via a Pachner 3$\rightarrow$2 move, it can not
because its dual is not homeomorphic to two tetrahedra simplices
joined at a face.}
\label{3duncontra}
\endfig

The above observations suggest that extra restrictions can be imposed on when one is permitted to do a
Pachner move. The restrictions one imposes depends on what
characteristics one is looking for in the theory. We will consider
four ``natural" rules: the Basic Rule, the Closed Rule,
a modified version of the Closed Rule (first introduced
by Smolin and Wan in \cite{SmoWan07}) and what we shall call the
Embedded Rule.

%%%%%%%%%%%%%%%%%%%%%%%%%%%% B A S I C %%%%%%%%%%%%%%%%%%%%%%%%%%%%%%%%%
\subsection{Basic Rule}
The Basic Rule consists simply of applying the Pachner moves without any
restrictions. More precisely, what we will call the Basic Rule is
the set of rules for which all Pachner moves are permitted by the dynamics. We will later consider dynamics where this is not the case and
Pachner moves are forbidden in certain situations. As such, the Basic Rule is the
least restrictive dynamical rule.
This rule, having the least restrictions, will have the least
number of quantities conserved by the evolution.
%%%%%%%%%%%%%%%%%%%%%%%%%%%% F O T I N I    R U L E %%%%%%%%%%%%%%%%%%%%%%%%%%%%%%%%%
\subsection{Closed Rule}
The Closed Rule is motivated by the idea that the microscopic evolution of space can only happen arround areas which are not convoluted and twisted but rather have trivial topology.
In classical General Relativity it is always the case that, locally, space has trivial topology.
As such, under Closed rule, the Pachner moves, which act on only a few simplices of space at a time, can only act if these simplices form a structure of trivial topology.  That is, the evolution is generated by
 Pachner moves but there is an extra restriction
as to when a Pachner move may be applied: the Pachner move may
only be applied when the simplices supporting the
 move are homeomorphic to a closed (each simplex is closed topologically) $n$-ball (a disc in 2d
and a ball in 3d).

%%%%%%%%%%%%%%%%%%%%%%%%%%%% W A N    R U L E %%%%%%%%%%%%%%%%%%%%%%%%%%%%%%%%%
\subsection{Smolin-Wan (Open) Rule }
In \cite{SmoWan07}, Smolin and Wan suggested a modification of
the Closed Rule. The same principle is applied as in the
Closed Rule except that the boundary of the union of
the tetrahedra which are acted upon by the Pachner move is disregarded. That is, one takes the
topological interior of the union of the simplices supporting the
Pachner move and if the resulting set is homeomorphic to an open
$n$-ball, the Pachner move is allowed, if not, the Pachner move is
forbidden. For example, in the configuration of the
two triangles on the top of \fg{fyrule}, according to the
Closed Rule, we cannot apply the \bb\  Pachner move
because the configuration is not homeomorphic to a closed disc, but
according to Smolin and Wan's Rule, the \bb\ move can be applied
because the interior is homeomorphic to an open disc.

\begin{figure}
\begin{center}
\scalebox{.7}{\includegraphics{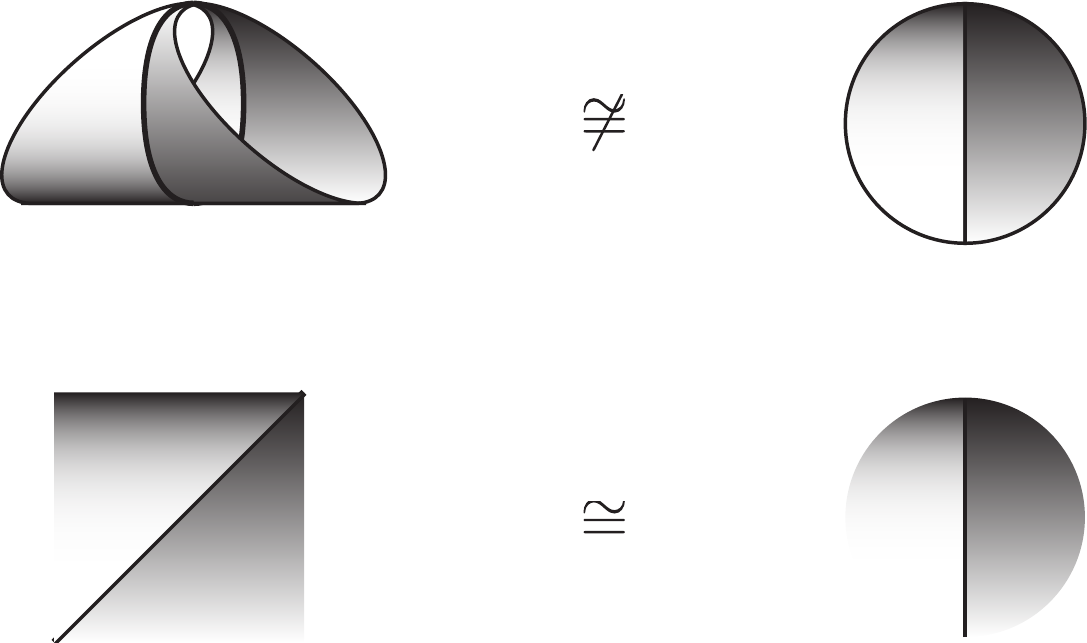}}
\end{center}
\capt{Due to the identification  of vertices on the grey triangle
and the white triangle, the resulting topology is not
homeomorphic to a closed disc.  The interior of the union of
the two triangles is homeomorphic to an open disc because it
excludes the graph vertices which are on the boundary.}
\label{fyrule}
\end{figure}

An alternative way (slightly less elegant but practical in actual
evaluations) to enforce Smolin \& Wan's Rule is simply to work with the
truncated graph. The truncated graph is the subgraph containing only
the nodes which take part in the Pachner move and the legs joined to
these nodes. By truncating the graph thus, we disregard any
relations between the exterior lines (the lines which have been
cut by the truncation because they are not part of a leg linking nodes
which participate in the Pachner move) that are dual to the boundary of the set of simplices
on which the Pachner move is applied. By doing that, we lose track of
possible identifications on the boundary which result from the
relations between the exterior lines. This contrasts with the Closed Rule where we must track how the exterior lines link up.
%%%%%%%%%%%%%%%%%%%%%%%%%%%%%%%%%%%%%%%%%%%%%%%%%%%%%%%%%%%%%%%%
%%%%%%%%%%%%%%%%%%%%%%%%%%%%EMBEDDED RULE%%%%%%%%%%%%%%%%%%%%%%%%%%%%
\subsection{Embedded Rule}
The three previous rules apply when the line
graph or framed graph is abstract graph, not a graph embedded in
a 3-manifold.  We can ask  whether these
Rules carry over to embedded line graphs or framed graphs. (Note that the we get different models for embedded framed graphs or an embedded line graphs: an embedded line graph has many more degrees of freedom and is much more complex.)

The answer is that, in the case of embedded graphs, a
further restriction must be added to the pre-existing restrictions.
The extra restriction is as follows: a Pachner move may be carried
out if and only if it can be implemented by continuously deforming
the framed graph. The reason we must impose this further
restriction is because otherwise the Pachner moves would be
ambiguous. This ambiguity arises from the fact that the topology of the graph (not that of the dual simplicial complex) can change during a Pachner move and so it may be impossible to embed the resulting graph in the same way the initial graph was embedded. This illustrated in \fg{embedproblem}.

To resolve the ambiguity, we impose
that the initial state must continuously deform into the final state. This uniquely defines the embedding of the final state.

\begin{figure}
\begin{center}
$a)$
\scalebox{.3}{
\includegraphics{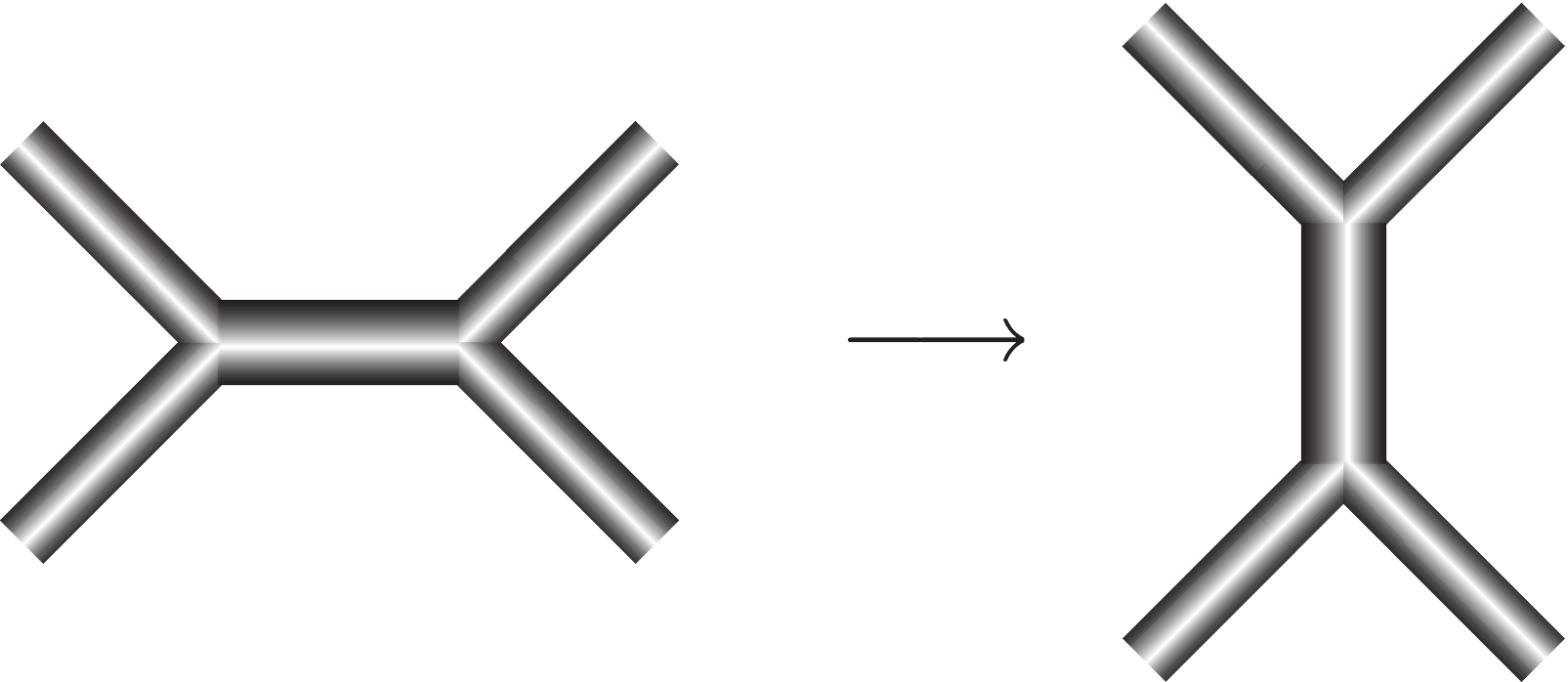}}
\qquad
$b)$
\scalebox{.3}{
\includegraphics{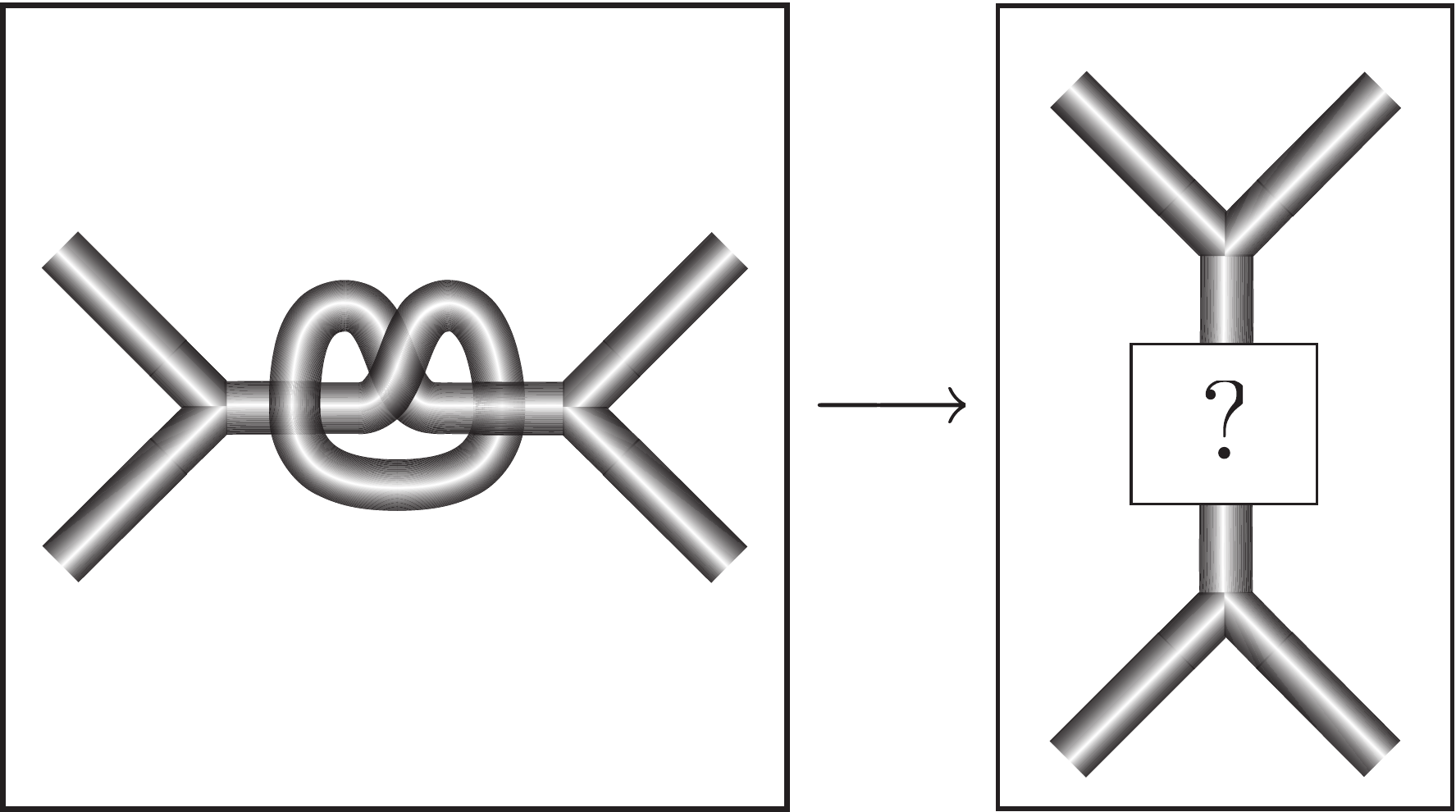}}
\end{center}
\capt{a) The standard \bb\ Pachner move in the non-embedded case. b) The embedded case is trickier: if we want to implement the \bb\ Pachner move on
the embedded framed graph containing a knot, how do we choose how to
embed the resulting framed graph? Should we put a right trefoil
knot, a left trefoil knot, or something else in the place of the
question mark on the right? It is ambiguous. }
\label{embedproblem}
\endfig

An {abstract} graph can always be ``continuously deformed" in the
sense that, for example, we can see the graph as being embedded in $\R^4$ where
all knots and braids can be undone. Without knots and braids, all Pachner moves can be carried out by continuously deforming the graph.

%%%%%%%%%%%%%%%%%%%%%%%%%%%%% RULE HIERarCHY %%%%%%%%%%%%%%
\subsection{Rule Hierarchy}
It is important to note that there is a hierarchy of Rules. The
Basic Rule has the least constraints. Smolin \& Wan's Rule has all
the constraints of the Basic Rules plus some extra ones.
Closed Rule has all the constraints of Smolin \&
Wan's Rule plus some extra ones. These three Rules can therefore be
ordered from the most constraining (Closed Rule) to
the least constraining (Basic Rule). These relations imply
similar relations in terms of the conserved quantities of each
dynamical model. The fact that Smolin \& Wan's Rule respects all the
constraints of the Basic Rule means that any quantity that is
conserved for the Basic rules is also conserved for Smolin \& Wan's
Rule. And similarly, the fact that the Closed Rule
respects all the constraints of Smolin \& Wan's Rule implies that
any quantity that is conserved for Smolin \& Wan's Rules will also
by conserved for the Closed Rule.

The Embedded Rule is independent from the previous three rules in
the sense that it can be applied in conjunction with any of the
three aforementioned rules and simply adds extra constraints and
therefore extra conserved quantities (infinitely many extra
conserved quantities in fact, due to all the possible knots which
are conserved) when the graph is an embedded one.

We now turn to the focus of this paper. It would be very nice to see whether matter degrees of freedom can be realized as topological defects in the 3d case, but
to help us see what we should be looking for and what to expect we
will first thoroughly analyze the 2d case.

%%%%%%%%%%%%%%%%%%%%%%%%%%%%%%%%%%%%%%%%%%%%%%%%%%%%%%%%%%%%%%%%%%%%%%%%%%%%%%%%%%%%%%%%%%%%%%%%%
%%%%%%%%%%%%%%%%%%%%%%%%%%%%%%%%%%%%%%%%%%%%%%%%%%%%%%%%%%%%%%%%%%%%%%%%%%%%%%%%%%%%%%%%%%%%%%%%%
%%%%%%%%%%%%%%%%%%%%%%%%%%%%%%%%% M A I N  S E C T I O N %%%%%%%%%%%%%%%%%%%%%%%%%%%%%%%%%%%%%%%%
%%%%%%%%%%%%%%%%%%%%%%%%%%%%%%%%%%%%%%%%%%%%%%%%%%%%%%%%%%%%%%%%%%%%%%%%%%%%%%%%%%%%%%%%%%%%%%%%%
%%%%%%%%%%%%%%%%%%%%%%%%%%%%%%%%%%%%%%%%%%%%%%%%%%%%%%%%%%%%%%%%%%%%%%%%%%%%%%%%%%%%%%%%%%%%%%%%%

%%%%%%%%%%%%%%%%%%%%%%%%%%%%%%%%%%%%%%%%%%%%%%%%%%%%%%%%%%%%%%%%%%%%%%%%%%%%%%%%%%%%%%%%%%%%%%%%%
%-----------2D-2D-2D-2D-2D-2D-2D-2D-2D-2D-2D-2D-2D-2D-2D-2D-2D-2D-2D-2D-2D-2D-2D-2D-2D-2D-2D-2D-%
%%%%%%%%%%%%%%%%%%%%%%%%%%%%%%%%%%%%%%%%%%%%%%%%%%%%%%%%%%%%%%%%%%%%%%%%%%%%%%%%%%%%%%%%%%%%%%%%%

%%%%%%%%%%%%%%%%%%%%%%%%%%%%%%%%%%%%%%%%%%%%%%%%%%%%%%%%%%%%%%%%%%%
\section{Conserved Quantities in the 2d Case}

We shall now turn to the conserved quantities under each of the above Rules.
As mentioned previously, the
Basic Rule has the least conserved quantities and each successive
Rule adds more conserved quantities. We will, therefore, first
consider the Basic Rule and then, for each other rule, find what conserved
quantities are added.
%%%%%%%%%%%%%%%%%%%%%%%% B A S I C    R U L E %%%%%%%%%%%%%%%%%%%%%%%%%%%%%%%BASIC
\subsection{The Basic Rule: Classifying Topology in 2d}
%\subsubsection{Global Topology}

The topology of connected 2D manifolds can be classified by whether
the manifold is orientable or not and a natural number. A good way
to see this is through prime decomposition which decomposes
manifolds into the connected sum of prime manifolds. The connected
sum,$M\#N$, of two n-manifolds, $M$ and $N$, is obtained by cutting out an
$n$-ball from each manifold and gluing the resulting boundaries
together (with the proper orientation if the manifolds are
orientable).  The result is unique up to homeomorphism. The connected
sum of two tori is illustrated in \fg{csum}.

\begfig
\begin{center}
\begin{equation}
\stackrel{
\includegraphics[width=0.1\vsize]{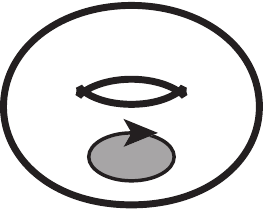}}
{M}
\quad
\stackrel{
\includegraphics[width=0.1\vsize]{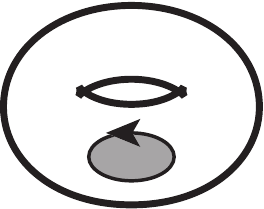}}
{N}
\qquad
\stackrel{
\includegraphics[width=0.2\vsize]{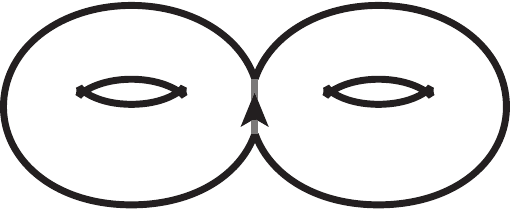}}
{M\#N}
\nonumber
\end{equation}
\end{center}
\capt{The connected sum of two tori.} \label{csum}
\endfig

The connected sum is obviously commutative and it is also
associative, the unit element is the $n$-sphere. A prime manifold is
a manifold which is not homeomorphic to any non-trivial connected
sum, i.e., it cannot be decomposed as the connected sum of two other manifolds other than the connected sum of itself with an $n$-ball. In 2d, there are only two prime
manifolds: the torus and the real projective plane (the projective plane is the 2-sphere with each point identified with its antipodal point, it is also the M\"obius strip with its edge contracted to a single point).  We can
therefore characterise the global topology of the manifold with two
numbers: the number of tori and the number of projective planes in
its prime decomposition\cite{topo}. The decomposition is not unique. As
we will see later, if there is at least one projective plane, then
each torus can be replaced by two projective planes without changing
the topology. If the prime decomposition does not contain any
projective plane, then the manifold is orientable and is
characterised by the number of tori in its prime decomposition, also
called its genus. If the prime decomposition contains a projective
plane, then the manifold is not orientable and it is homeomorphic to
a manifold with a prime decomposition containing only projective
planes. Therefore, a non-orientable manifold $M$ is characterised by
the number of projective planes it takes in a connected sum without
tori to build a manifold homeomorphic to $M$.

To recap, the topology of connected 2d manifolds can be categorised by whether the
manifold is orientable or not and a natural number. If the manifold
is orientable, the natural number represents its genus (the number
of holes or ``handles" the manifold contains) and, in the case of a
non-orientable surface, the natural number represents the number of
M\"obius strips (also called cross-caps) glued to the boundary of a
disc. To add a cross-cap to a surface, one has only to make a slight
cut on the surface and then reglue the two sides together with opposite orientation as shown in \fg{surgery}.

\begin{figure}
\begin{center}
\includegraphics[width=0.45\hsize,angle=0]{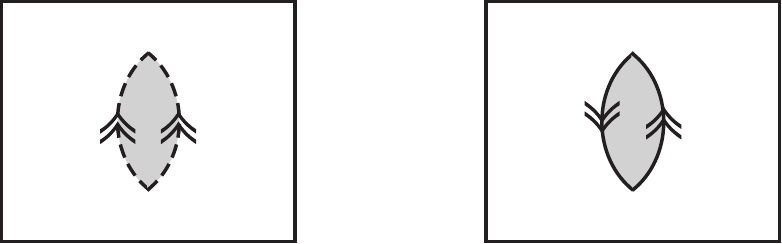}
\end{center}
\capt{Operation of adding a cross-cap to a surface. A
piece of the oriented surface is cut out, flipped orientation, and
glued back in.  As a result, the surface has acquired a cross-cap. } \label{surgery}
\endfig

Adding two cross-caps we get a ``Klein bottle handle" which can also
be represented by cutting out two circles and regluing them with the
same orientation. As shown in \fg{tor}, this is very similar to the
to a normal handle except that, for a normal handle, the two circles
which are glued back together have opposite orientation.

\begfig
\begin{center}
\begin{equation}
a)\ \
\includegraphics[height=0.2\vsize]{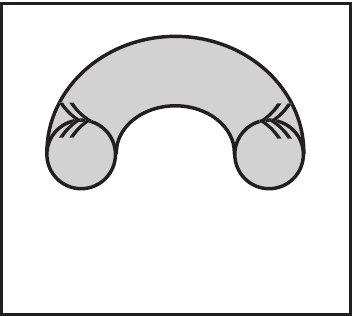}
\qquad
b)\ \
\includegraphics[height=0.2\vsize]{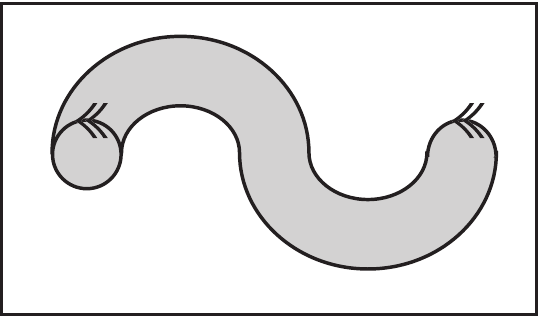}
\nonumber
\end{equation}
\end{center}
\capt{a) Gluing a normal handle on a surface. b) Gluing a
Klein bottle handle on a surface. For the normal handle, the
two circles have opposite orientation whereas for the Klein bottle
handle,  the circles have the same orientation.  } \label{tor}
\endfig

Since, as mentioned in Section \ref{SectionDynamics}, the Pachner moves
allow the transition from any initial state to any final state
\emph{as long as the initial state and the final state have the same
topology}, we must conclude that the only conserved quantity for the
Basic rule is the global topology. In other words, the conserved
quantity is $Q_B \equiv 2 n_T + n_{PP} $ and the orientation of the
manifold, where $n_T$ is the number of tori in a prime decomposition
of the manifold and $n_{PP}$ is the number of projective planes in the
same decomposition.

In a theory based on such graph evolution, it may be
 interesting to consider the handles and
cross-caps to be pseudo-particles. We say pseudo-particles because even though they can propagate, diffuse and interact as we will see, it is not clear how they relate to real particles in some semiclassical background.
In any case,
the handles and the cross-caps  are not always precisely located and can
spread across arbitrarly large distances.  This could be a problem if we expect that a handle
spread over macroscopic distances and make space topologically non-trivial on observable scales.

Understanding the significance of these pseudo-particles requires
 being able to
recognise them in our graph. This is not so obvious. In \fg{2dparts}
of Appendix C we can see three configurations which corresponded to (a)
a disk, (b) \& (c) a cross-cap and (d) a handle. The situation is
much more complicated in general as, in order to figure out, for
example, what topology the joining of subgraph of \fg{2dparts} (b) with the
subgraph of \fg{2dparts} (d) corresponds to, we need to
manipulate the graph until we get a familiar shape.

One way to start reading the topology off the graphs is to recognize that, in
addition to what we already know about the subgraphs in
\fg{2dparts}, the subgraphs presented in \fg{isleofman} correspond
to (if embedded in a larger graph such that the three lines of
each subgraph form three different loops, i.e., that the subgraph
represents a triangle with three distinct vertices): (a) adding a
disc to the larger graph, i.e. not changing the topology of the
larger graph, (b) adding a cross-cap to the larger graph, (c) adding
2 cross-caps to the larger graph, and (d) adding a handle to the
larger graph.  Adding a disc means taking the connected sum with a 2-sphere (this
leaves the original manifold invariant), adding a handle refers to
taking the connected sum with a torus, as shown in \fg{csum}, and
adding a cross-cap refers to taking the connected sum with a
projective sphere. Adding a cross-cap is equivalent to cutting out a disc from
the original manifold and gluing the edge
of a M\"obius strip on the edge of the hole, or equivalently, to performing the cut of \fg{surgery} on the
manifold.
Inserting \fg{isleofman} in a larger graph such that each line of
\fg{isleofman} closes into a different loop adds a cross-cap to the
larger graph as shown in \fg{tricap}.

\begin{figure}
\begin{center}
\begin{equation}
\begin{array}{rcrc}
a)&\
\scalebox{1}{\includegraphics{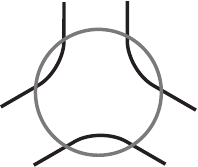}}&
\qquad b)&
\scalebox{1}{\includegraphics{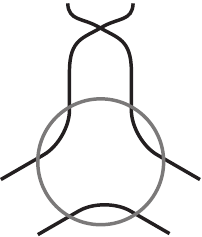}}\\
&&&\\
c)&
\scalebox{1}{\includegraphics{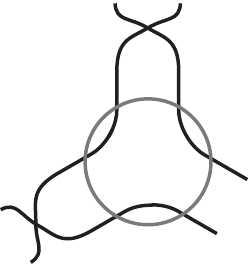}}&
\qquad d)&
\ \scalebox{1}{\includegraphics{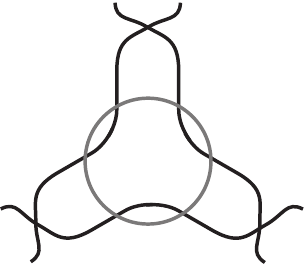}}
\end{array}
\nonumber
\end{equation}
\end{center}
\capt{For each subgraph in this figure, if it is embedded in a
larger graph such that the three lines of the subgraph close into
three different loops then, (a) adds a disc to the larger graph (i.e.
does not change the topology of the larger graph), (b) adds a
cross-cap to the larger graph, (c) adds 2 cross-caps to the larger
graph and (d) adds a handle to the larger graph. } \label{isleofman}
\endfig

\begin{figure}
\begin{center}
\scalebox{1}{
\includegraphics{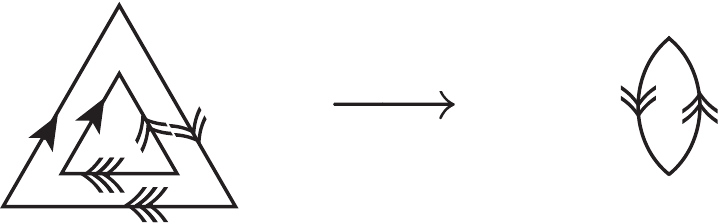}}
\end{center}
\capt{Inserting the subgraph of \fg{isleofman} (b) in a larger graph
corresponds, topologically, to adding a cross-cap to it.}
\label{tricap}
\endfig

Therefore, to figure out the dual topology, we gradually remove all subgraphs
of the type shown in \fg{isleofman}, counting how many cross-caps and
handles have been removed, until we are left with a graph with at
most three loops, at which point we use the Pachner moves to get to a
configuration of where we know the topology.

%%%%%%%%%%%%%%%%%%%%%%%%%%%% W A N %%%%%%%%%%%%%%%%%%%%%%%%%%%%%%WAN
\subsection{Smolin \& Wan's Rule}

As mentioned previously, because Smolin \& Wan's Rule is the Basic
Rule with added restrictions, we already know that the conserved
quantities of the Basic Rule are still conserved with Smolin \&
Wan's Rule. Therefore, the number of handles and cross-caps is also
fixed in Smolin \& Wan's Rule. The new restriction brought about in
the evolution by Smolin \& Wan's rule is that a $j\rightarrow k$
Pachner move can now only be carried out if the subgraph
containing only the $j$ nodes on which the Pachner move is to be
applied is homeomorphic to an open $n$-ball. In 2d we
have three possible Pachner moves: \ca , \ac\ and \bb . We will check
these moves one by one to see what restrictions Smolin \& Wan's Rule
adds.

\ca : To perform the \ca\ Pachner move under the Basic rule, one must
start with the subgraph on the left in \fg{wan31}(a). When the free lines of the subgraph on the left in
\fg{wan31}(a) are connected to nodes other than the three nodes in
the subgraph, Smolin \& Wan's Rule allows for the \ca\ Pachner
move to be performed. This is because the
subgraph containing the three nodes is then dual to a triangle
homeomorphic to a open disc. The only time the \ca\ move could possibly be forbidden
is when the free lines are directly connected to each other. There are
only two ways to link them to each other, of which only one,
presented on the left in \fg{wan31}(b), makes the dual not
homeomorphic to a open disc (it is homeomorphic to a M\"obius strip).
Therefore, with respect to the \ca\ move, Smolin \& Wan's Rule gives
a single further restriction with respect to the Basic move: the
move presented in \fg{wan31}(b) is forbidden with Smolin \& Wan's
Rule whereas it was not with the Basic Rule.

\begin{figure}
\begin{center}
\begin{equation}
a)
\scalebox{.8}{
\includegraphics{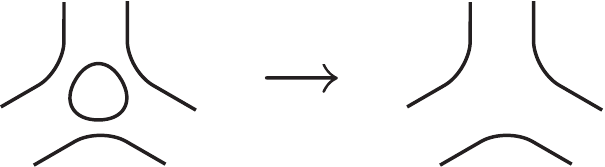}}\qquad
b)
\scalebox{.8}{
\includegraphics{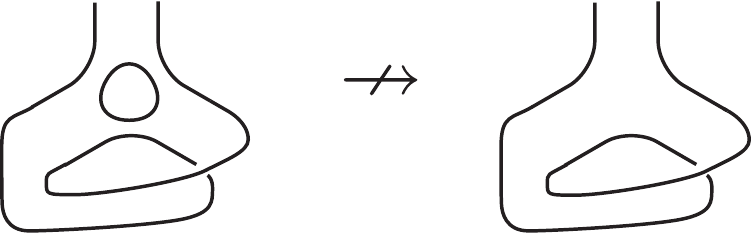}}
\nonumber
\end{equation}
\end{center}
\capt{With the Basic Rule, to perform the \ca\ Pachner move, one need
simply find a subgraph of the form in (a).  With Smolin \& Wan's
Rule, if two of the legs of the subgraph are directly joined
together as in (b), then the \ca\  Pachner move is forbidden.}
\label{wan31}
\endfig

\ac: To perform the \ac\  Pachner move, one need simply start with a
node. If the three free lines of the node are all linked to other
nodes, then Smolin \& Wan's move will not restrict the application
of the \ac\  move as the subgraph containing only the the node will be
dual to a triangle homeomorphic to a open disc. For possible restrictions
coming from Smolin \& Wan's Rule one must therefore have the links
between the three lines. There are only two possible ways to link
the the lines, those presented \fg{simplsimpart}. The
subgraph in \fg{simplsimpart}(a) is dual to a triangle with two of
its edges glued together to from a cone, which is homeomorphic to a
open disc and therefore Smolin \& Wan's Rule does not restrict such a
node from being expanded via the \ac\ move. The subgraph presented in
\fg{simplsimpart}(b), on the other hand, is dual to a triangle with
two edges identified so as to form a M\"obius strip (as shown in
\fg{trimoebius}) and so the \ac\ move which expands such a subgraph
is forbidden. This restricted Pachner move is the reverse of the
move in \fg{wan31}(b).

\begin{figure}
\begin{center}
$a)$
\scalebox{1}{\includegraphics{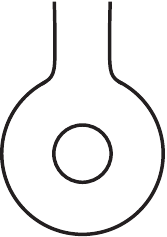}}
$\qquad b)$
\scalebox{1}{\includegraphics{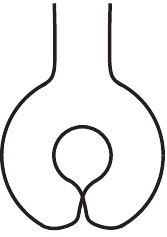}}
\end{center}
\capt{The only two possible ways to have non-trivial topology with
only one simplex: a) This graph is dual to a triangle with two of
its edges glued together to from a cone, which is homeomorphic to an
open disc. b) This graph on the other hand is dual to a triangle with
two edges identified so as to form a M\"obius strip.} \label{simplsimpart}

\endfig
\begin{figure}
\begin{center}
\scalebox{1}{
\includegraphics{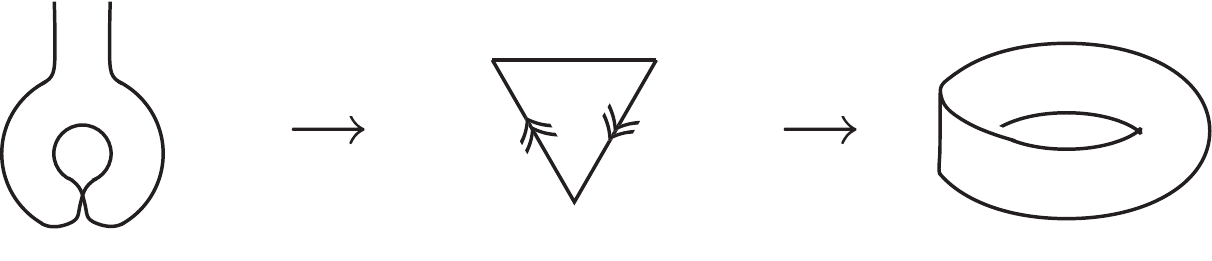}}
\end{center}
\capt{The dual of the subgraph on the left
is a M\"obius strip.} \label{trimoebius}
\endfig

\begin{figure}
\begin{center}
\scalebox{1}{
\includegraphics{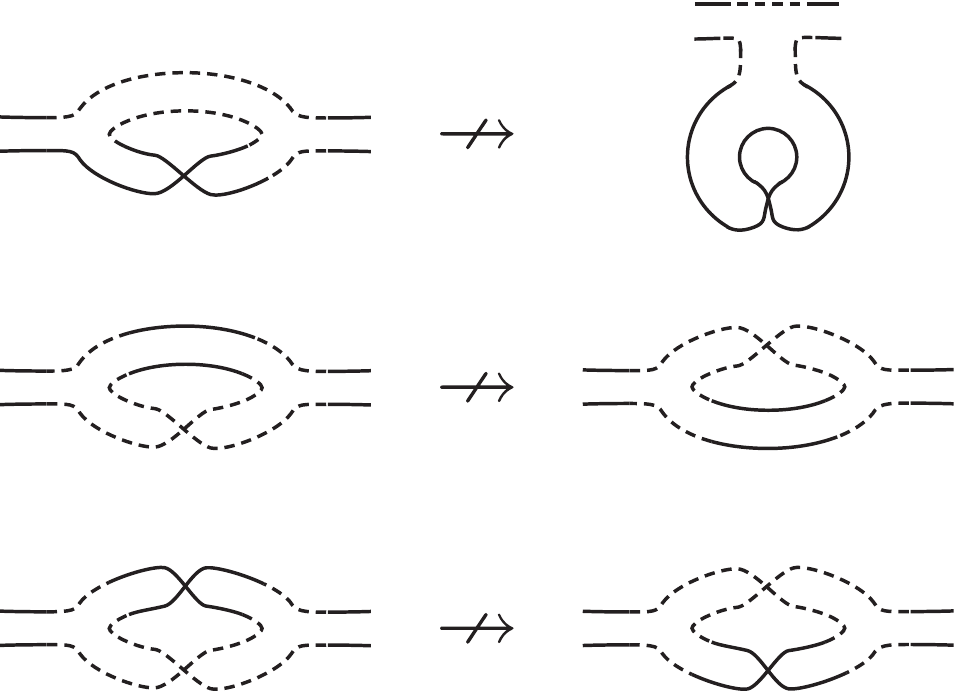}}
\end{center}
\capt{The leg along which the \bb\ move is performed is marked by a dotted line.}
\label{wan22}
\endfig

\bb: Following the same procedure, one finds that the \bb\ Pachner
moves which are forbidden by Smolin \& Wan's Rule are the ones
presented in \fg{wan22}. Note that 2-node-subgraphs with leg
crossings are equivalent to 2-node-subgraphs without leg crossings
but with extra line crossings (see Appendix C).  This is why leg
crossings are not considered below.
Of these, the move in \fg{wan22}(c) does nothing, and so,
even though forbidding it with Smolin \& Wan's Rule can potentially change the
dynamics by modifying transition amplitudes, forbidding this move
has strictly no effect on conserved quantities.  A similar situation
occurs for the move in \fg{wan22}(b). In that case the move does in
fact change the graph, but a sequence of legitimate moves (under
Smolin \& Wan's Rule) can achieve the same result as shown in
\fg{wananex} of Appendix B. Hence, the fact that this move is
forbidden under Smolin \& Wan's Rule has no influence on the
conserved quantities. %, only the transition amplitudes might be modified.

Therefore, the only restrictions that will add new conserved quantities
to those of the Basic rule are the Pachner move
shown in \fg{wan31}(b) and its reverse and the Pachner move shown in
\fg{wan22}(a) and its reverse. These moves are precisely all the
moves which create or destroy the subgraphs of
\fg{simplsimpart}(b) which correspond to a single simplex M\"obius
strip (cross-cap).

We have then found that  the quantities conserved by Smolin \&
Wan's Rule are the global topology (the number of handles and
cross-caps as in the Basic
Rule) and the number of simple simplex M\"obius strips dual to
\fg{simplsimpart}(b).

%%%%%%%%%%%%%%%%%% M A R K O P O U L O U - K A L A M A R A %%%%%%%%%%%FOTINI
\subsection{Closed Rule}

The Closed Rule, which forbids Pachner
moves when the closure of their support is not homeomorphic to a
closed disc, is more restrictive than Smolin \& Wan's Rule. Passing
from Smolin \& Wan's Rule to the Closed Rule, one adds a
countably infinite number of conserved quantities. For each $n\in\N$
there is a finite number of configurations involving $n$ simplices
which are conserved by the evolution. This is quite different from
Smolin \& Wan's Rule in which only configurations, involving a single
simplex are conserved. To see that for every $n$ there is at least one
configuration that is conserved by the evolution, we look at
\fg{fotinfinity} whose dual consists of $n=2k$ simplices (or $n=2k+1$
simplices if we include the node in brackets), none of which are
homeomorphic to a closed disc. All these simplices are glued together and
cannot be separated or modified by the evolution because any choice
of a subset of those simplices is not homeomorphic to a closed disc. One
can equally see that, for every $n$, there is only a finite number of
such configurations because there is only a finite number of ways to
glue $n$ simplices together.  For this reason, we will only
investigate such conserved quantities made up of only up to two
simplices.

\begin{figure}
\begin{center}
\scalebox{.9}{
\includegraphics{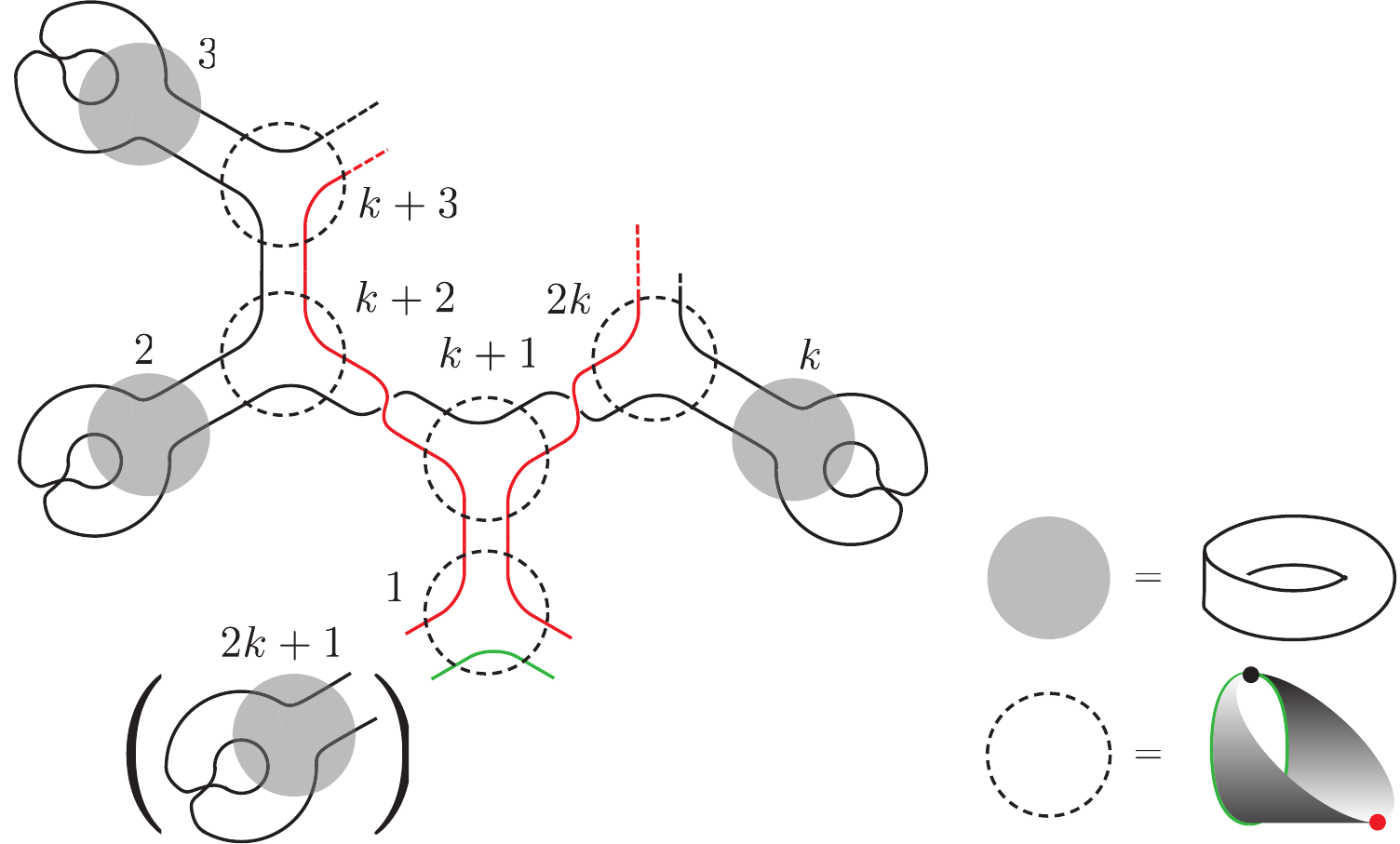}}
\end{center}
\capt{This configuration of $n$ nodes (where $n = 2k$ and the node in
brackets is not included if $n$ is even and $n = 2k+1$ and the node in
brackets is included if $n$ is odd) is stable under
the Closed Rule because none of the simplices are
homeomorphic to a closed disc. Notice that this configuration is not stable
under Smolin \& Wan's Rule because the nodes marked by the dotted
circles can be expanded under Smolin \& Wan's Rule as the interior of their dual
is homeomorphic to an open disc.} \label{fotinfinity}
\endfig

\subsubsection{Conserved Quantities up to 2 Simplices}

A stable (with respect to the Closed Rule) configuration
of two simplices can either have the two simplices linked by one leg
or two legs. They cannot be linked by all three legs, otherwise there
will not be any free legs to link with the rest of the graph. Also, an
interesting phenomenon happens with the Closed Rule: a
subgraph can look locally (i.e., with respect to the graph connectivity) as if it is
dual to simplices which are homeomorphic to to a closed disc while,
in fact, looking at the whole graph, this is not that case (see \fg{intconfig}).

\begin{figure}
\begin{center}
\scalebox{1}{
\includegraphics{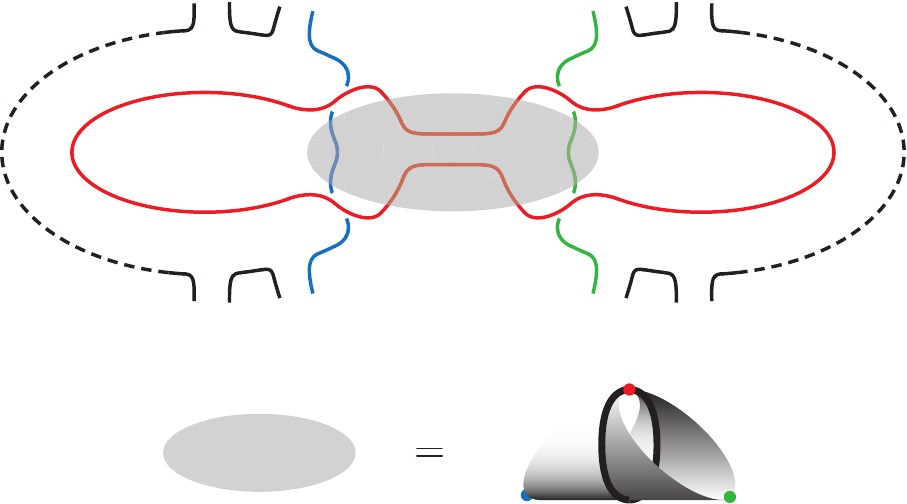}}
\end{center}
\capt{Taken out of context, the subgraph inside the shaded area
seems to be dual to two triangles joined along one edge forming a
parallelogram because the two red line segments seem to be part of
different lines. If that were the case, Pachner moves could be
applied to these nodes in accordance with the Closed
Rule. We need to look at the whole graph to see that the two red
line segments are part of the same line and that therefore no
Pachner moves can be applied to the nodes inside the shaded areas. }
\label{intconfig}
\endfig

For a specific configuration to be stable under the Closed Rule, \emph{no} single simplex that composes the stable substructure can be homeomorphic to a closed disk. This is because, if one of the simplices is homeomorphic to a closed disk, we can expand it to three simplices via the \ac\ Pachner move. As shown in \fg{fotonic1} here are only three ways for a simplex to fail to be homeomorphic to a disk: 2 vertices of the triangle are identified, three vertices of the triangle are identified or two of the edges are identified, so as to form a M\"obius strip (the only other way they could be identified forms a cone which is homeorphic to a disk).

\begin{figure}
\begin{center}
\begin{equation}
\begin{array}{cc}
{\rm A(a)}&
\scalebox{1}{
\includegraphics{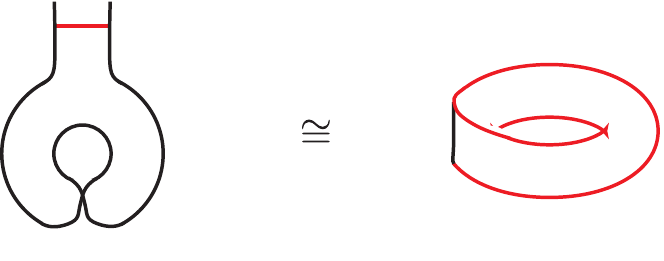}}
\\
\\
{\rm A(b)}&
\scalebox{1}{
\includegraphics{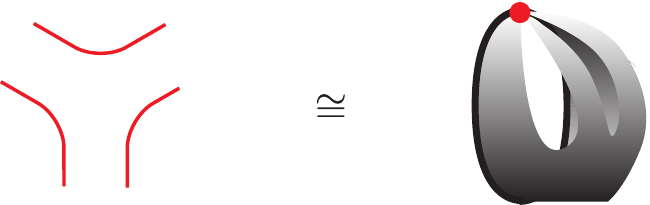}}\\
\\
{\rm B}&
\scalebox{1}{
\includegraphics{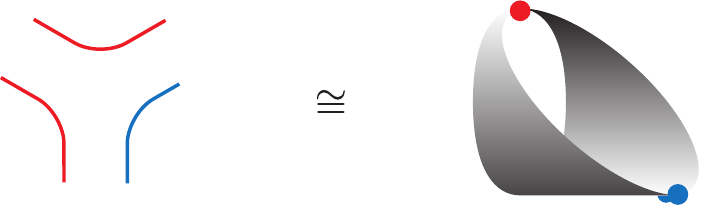}}
\end{array}
\nonumber
\end{equation}
\end{center}
\capt{The one simplex conserved configurations under the Closed rule. On the left is the line graph depiction of the conserved configurations, with similarly coloured line segments belonging to the same line. On the right is the simplicial complex dual to the line graph with the points having the same colour as their dual lines. Coloured bars across legs of the line graph indicate the colour of the dual edge in the simplicial depiction. In Category A are the conserved configurations which are conserved independently of their environment. In Category B are configurations which are conserved only in certain environments. Each of the above configurations give rise to a conserved quantity: the number of times they appear in a simplicial complex.} \label{fotonic1}
\endfig

The only possible stable structures under the Closed Rule made of a single triangle are therefore the ones shown in \fg{fotonic1}, but the fact that they cannot be expanded via a \ac\ move does not guarantee their stability.  To garantee their stability, we must make sure that no matter how it is connected to other triangles, no \bb\ or \ca\ Pachner moves can be applied to a set of triangles containing one of the structures of \fg{fotonic1}.

It is easy to see that this is in fact the case for \fg{fotonic1} A(a) because the simplex forms a non-orientable surface and so, no matter how it is joined to other simplices, any set of simplices containing this non-orientable simplex will itself be non-orientable and thus not homeomorphic to a closed disc. Thus, no Pachner move can be applied to a set of simplices containing the simplex of \fg{fotonic1} A(a). Therefore, \fg{fotonic1} A(a) is a conserved quantity under the Closed Rule. Of course, we could have deduced this from the fact that \fg{fotonic1} A(a) is the same as \fg{simplsimpart}(b) which we already saw is conserved under Smolin \& Wan's Rule and therefore must be conserved under Closed Rule because of the rule hierarchy.

\fg{fotonic1} A(b) is also conserved under the Closed rule. This is not as easy to see as in the case of \fg{fotonic1} A(a) but it follows from the fact that, to do a \bb\ move on \fg{fotonic1} A(b), it must be connected to a triangle that ``fills in" two of its holes. The only way to do that is to have a simplicial structure of the type shown in \fg{fotonic2} A(ii) which is not orientable and as such not homeomorphic to a closed disc. Furthermore, to perform a \ca\ on a subset of triangles containing \fg{fotonic1} A(b), at least one of the line segments of the line graph of \fg{fotonic1} A(b) must form a closed loop going through exactly three nodes. But this is impossible because all the line segments are part of the same line, and so the minimum number of nodes that line must go through before forming a closed loop is four.

It is not true in general, however, that the simplex configurations shown in Category B of \fg{fotonic1} will be invariant under the Closed Rule. In fact one can see that, if one of the holes of \fg{fotonic1} B(a) is ``filled in" with a simplicial complex homeomorphic to a closed disc, then the resulting simplicial complex will itself be homeomorphic to a closed disc. For example, suppose the two edges of the rim of the 2-triangle cone of \fg{2dcone} (a) of Appendix C were glued to the two edges bordering one of the hole of \fg{fotonic1} B(a).  The resulting simplicial complex is homeomorphic to a closed disc and, under the Closed rule, can be acted upon by the \ca\ Pachner move to get the simplicial dual to \fg{simplsimpart}(a).

On the other hand, if the free edges of the simplicial configuration of Category B are glued to a simplicial complex which either i) has non-trivial topology (i.e. is not homeomorphic to a closed disc) or ii) contains at least one configuration of Category A, then the simplicial configuration of Category B will necessarily be conserved by the Closed rule because its ``holes" will not be ``filled in" with simplices which can take part in a Pachner move with it.

Therefore, a single simplex configuration of Category A of \fg{fotonic1} is exactly conserved under the Closed rule independently of the greater simplicial complex it is in; but for a simplicial complex of Category B, whether it is conserved or not depends on its environment. We therefore define two categories of locally conserved simplicial configurations under the Closed rule: Category A simplicial configurations which are conserved independently of their environment and Category B simplicial configurations which are conserved only in certain environments. The overall number of each conserved simplicial configuration in a given simplicial complex defines a conserved quantity. A Category A quantity (the number of a specific Category A simplicial configuration in a simplicial complex) is always a conserved quantity. A Category B quantity is conserved only if the simplicial complex has non-trivial topology or contains at least one Category A conserved configuration.

Thus, we have seen that, up to one simplex, the conserved quantities of the
Closed Rule are the ones shown in \fg{fotonic1}. Similarly, the conserved quantities of the Closed Rule
made up of two simplices are the ones shown in \fg{fotonic2} and \fg{fotonic2b}.  Note that,
apart from \fg{fotonic2} A(iii), the two-simplex conserved configurations which are part of Category A are easily identifiable because they are either non-orientable or made up entirely from one-simplex Category A conserved configurations.

\begin{figure}
\begin{center}
\scalebox{.7}{
\includegraphics{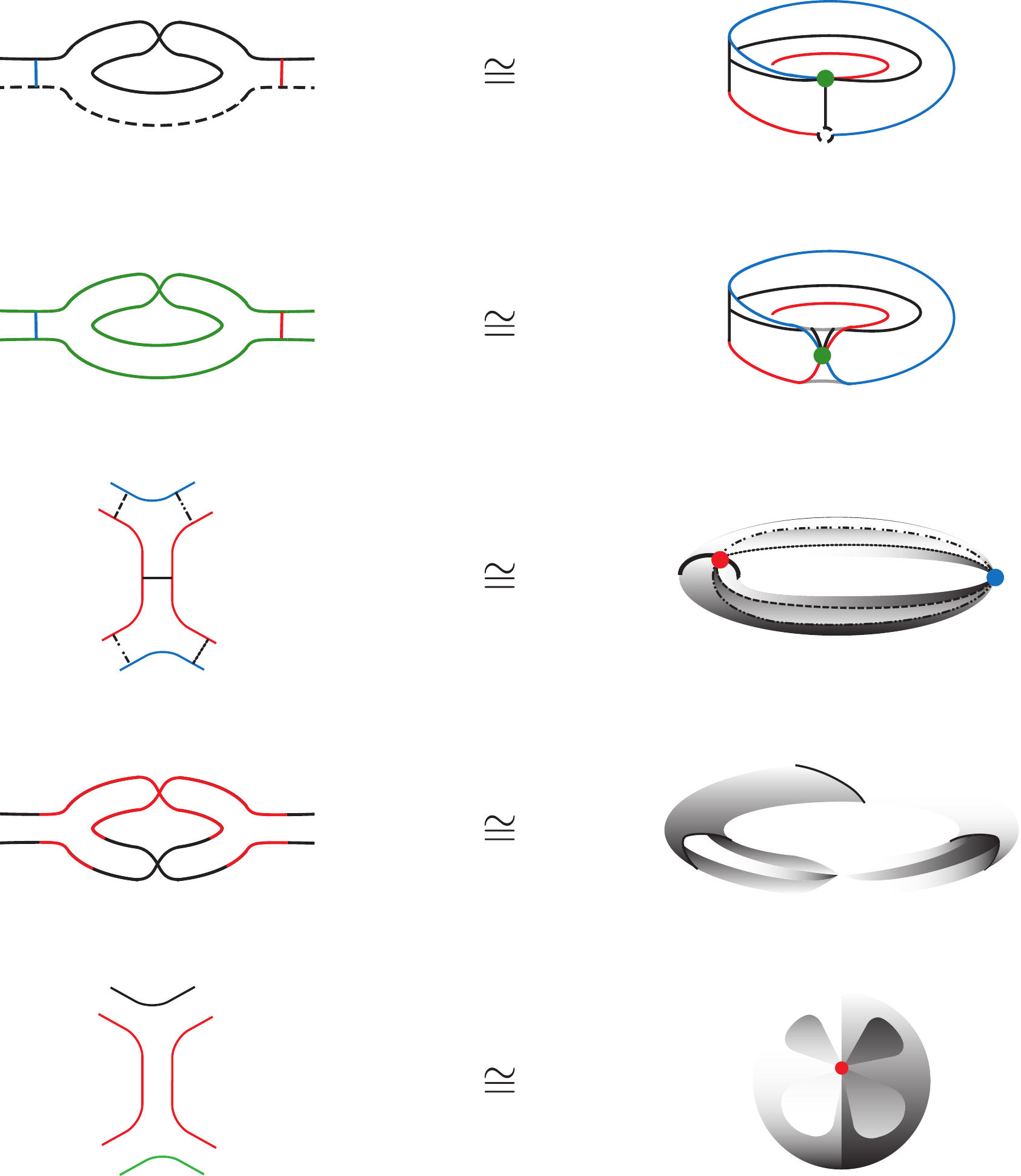}}
\end{center}
\capt{Two-simplex configurations conserved  under the Closed rule. On the left is the line graph depiction of the conserved configurations, with line segments of the same colour belonging to the same line. On the right is the simplicial complex dual to the line graph with the points having the same colour as their dual lines. Coloured bars accross legs of the line graph indicate the colour of the dual edge in the simplicial depiction. These are the conserved configurations which are conserved independently of their environment, i.e., Category A configurations. Each of the above configurations give rise to a conserved quantity: the number of times they appear in a simplicial complex.} \label{fotonic2}
\endfig

\begin{figure}
\begin{center}
\scalebox{.7}{
\includegraphics{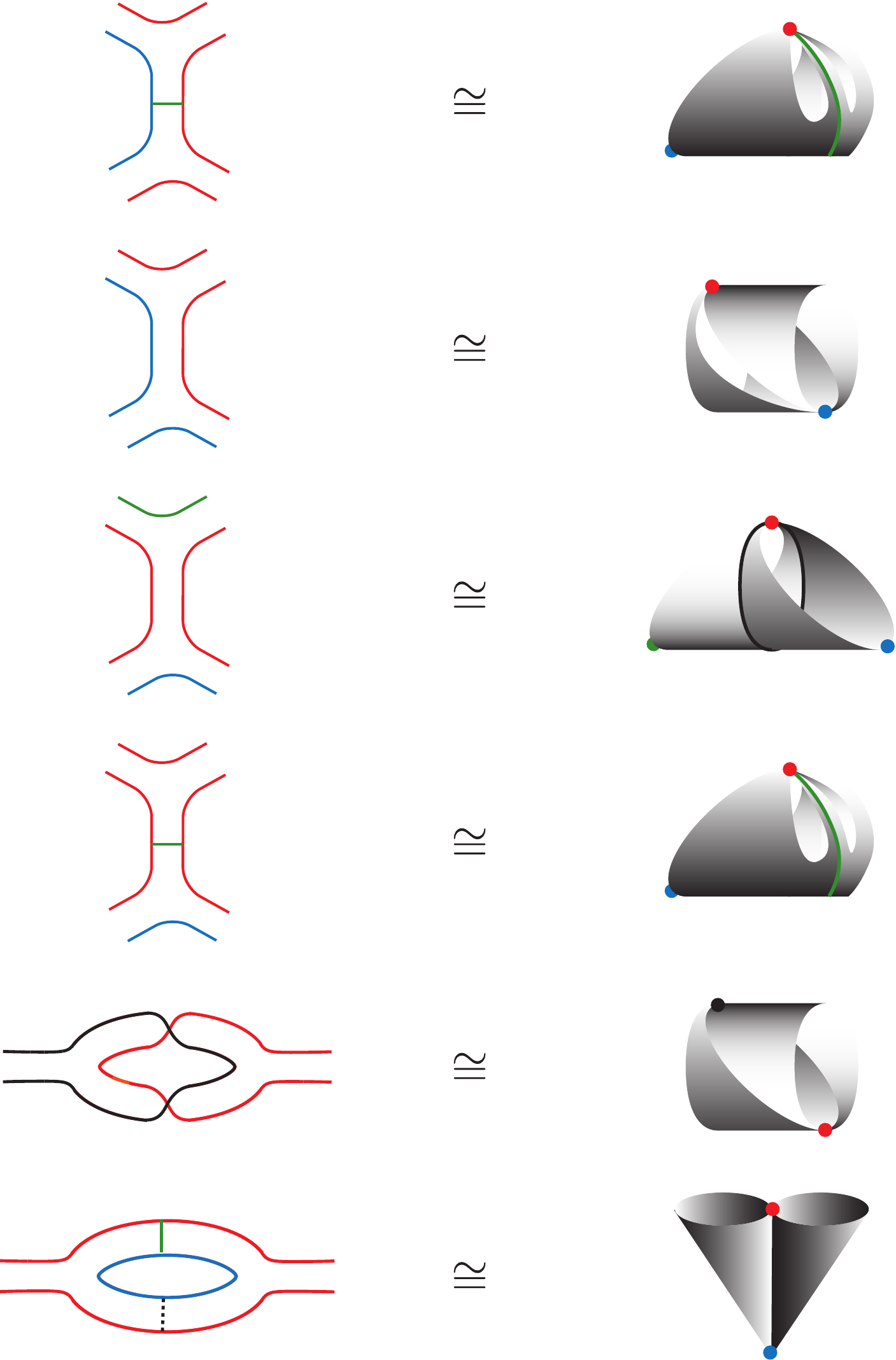}}
\end{center}
\capt{Similarly to the previous figure, these are the Category B configurations which are conserved  under the Closed Rule only in certain environments.} \label{fotonic2b}
\endfig

%%%%%%%%%%%%%%%%%%% E M B E D D E D  R U L E %%%%%%%%%%%%%%%%%%%%%%%%%%%%%%%EMBEDDED
\subsection{Embedded Rule}
There is, in fact, not one embedded rule but many. All the previously mentioned rules can be implemented in the embedded case, where the line graph or the framed graph is seen as embedded in a 3-manifold. The rules that were applied in the non-embedded case are still valid in the embedded case, but as explained in section 3.4, a new rule is necessary for the graph to be continuously deformed from the initial state to the final state inside the 3-manifold. This means that, in the embedded case, a specific model will have all the same conserved quantities as in the non-embedded case but will also gain an infinite number of new conserved quantities due to the braiding and knotting which are now possible and can restrict  move evolution as shown in \fg{embno}.
\begfig
\begin{center}
\begin{equation}
\begin{array}{lc}
a)&
\scalebox{.3}{
\includegraphics{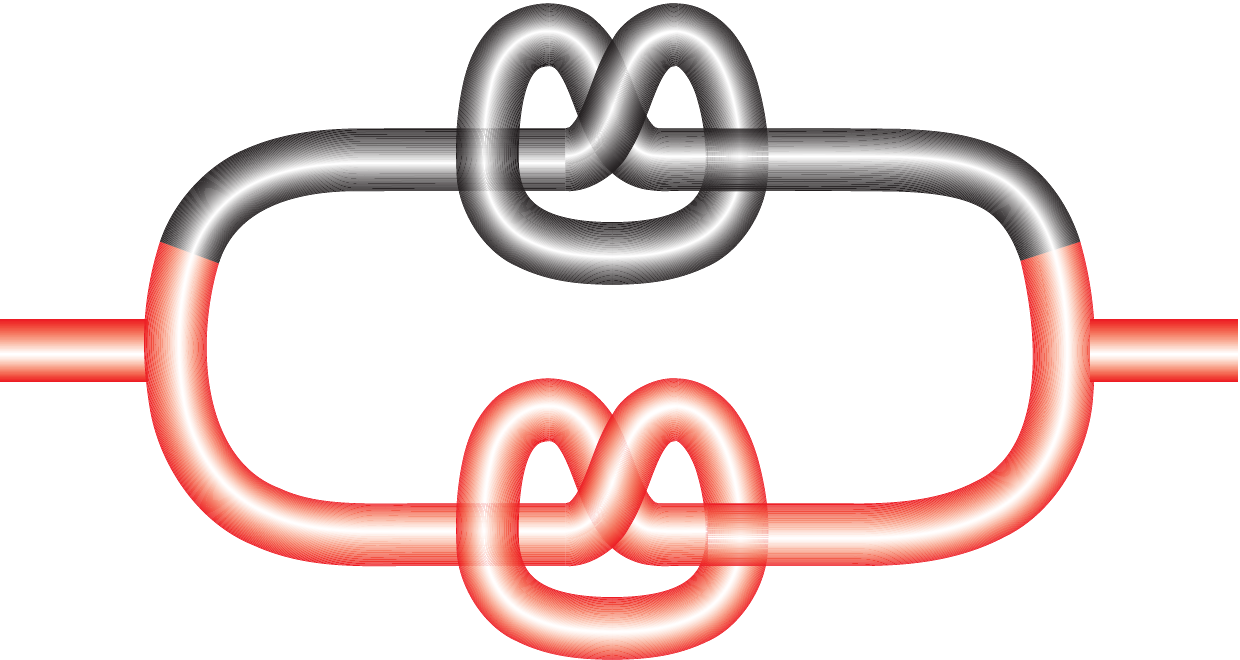}}\\
b)&
\scalebox{.3}{
\includegraphics{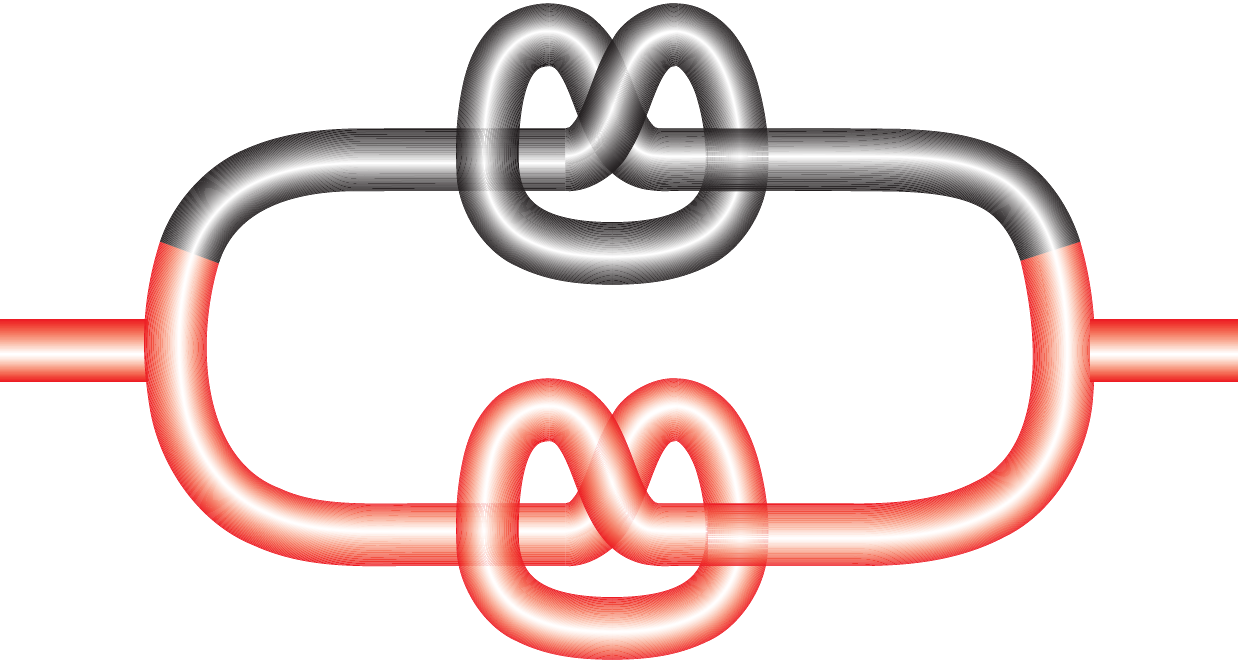}}\\
c)&
\scalebox{.3}{
\includegraphics{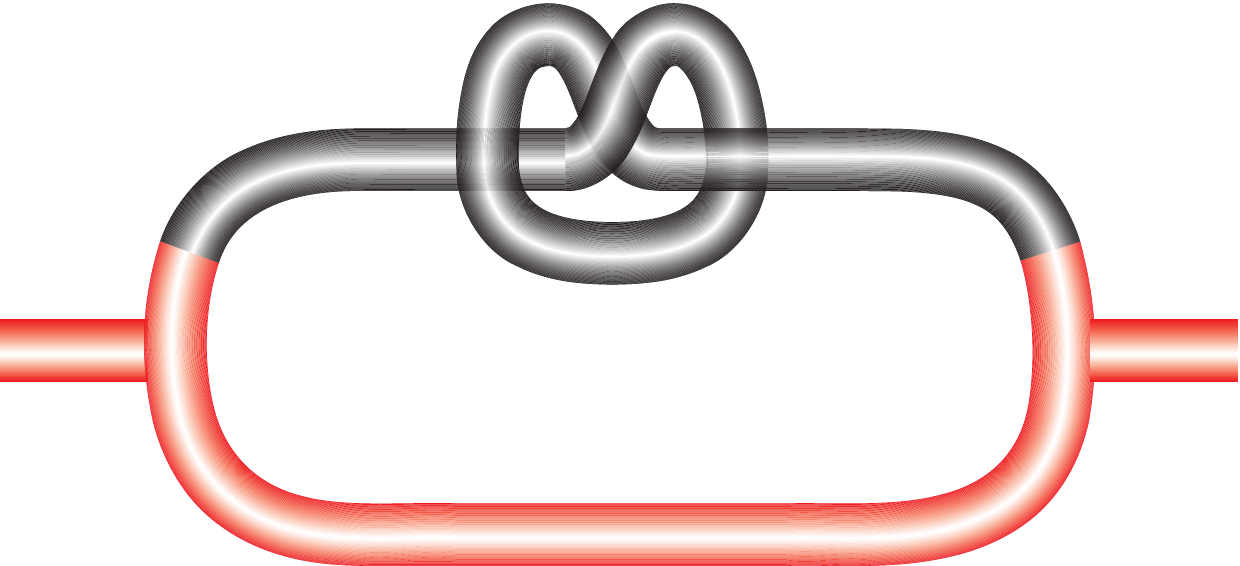}}\\
d)&
\scalebox{.3}{
\includegraphics{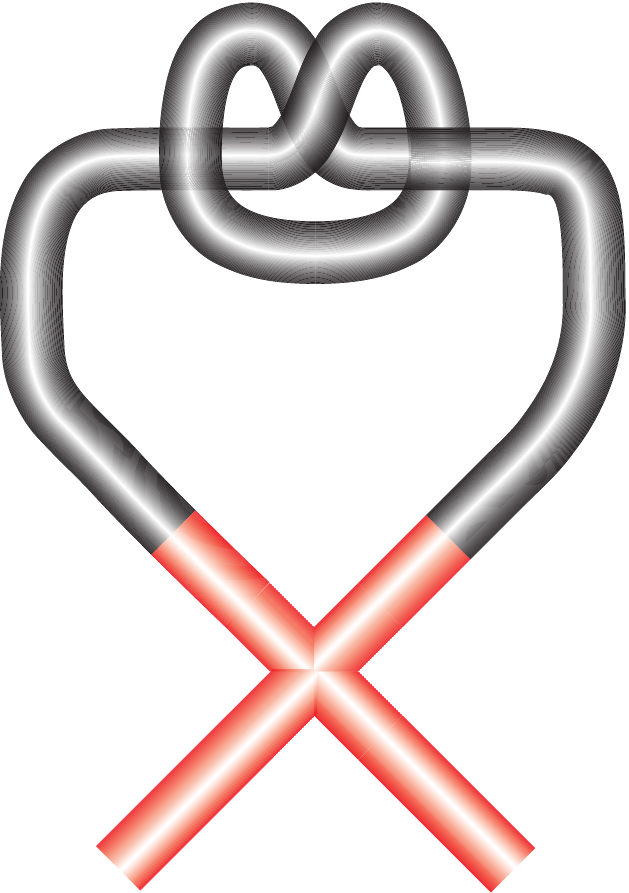}}\\
e)&
\scalebox{.3}{
\includegraphics{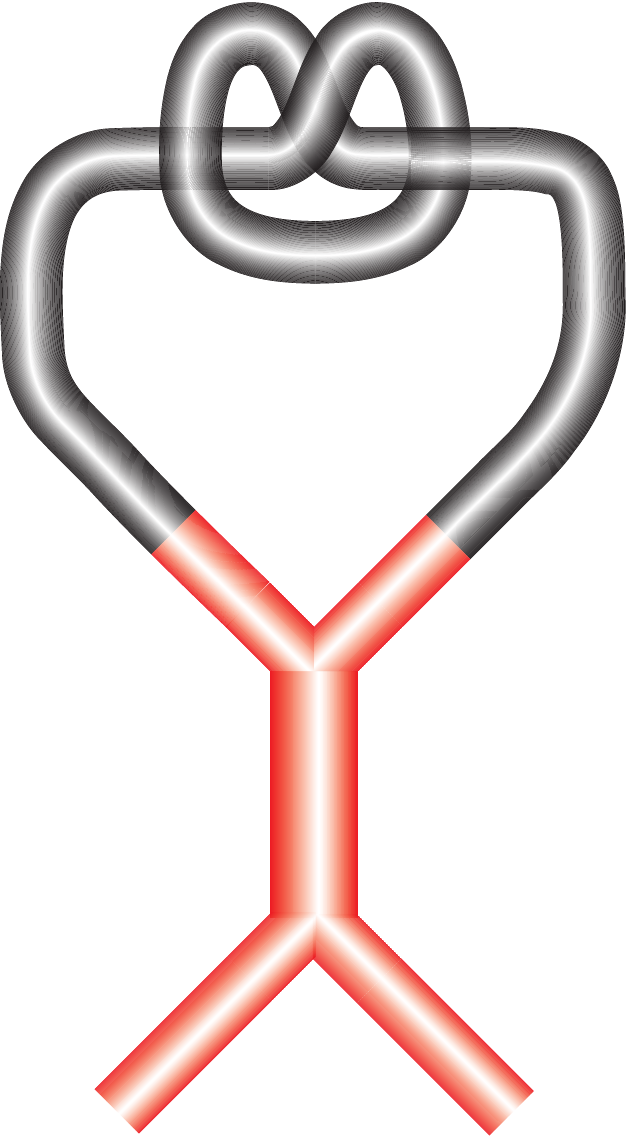}}
\end{array}
\nonumber
\end{equation}
\end{center}
\capt{If the framed graph is not embedded, there is a problem in
performing the 2$\rightarrow$2 Pachner move because whether the
crossing is over or under is irrelevant and so we can pass from (a) to
(b) which ultimately permits the crucial (c) to (d) step of contracting
the joining leg. On the other hand, if the framed graph is embedded,
we cannot continously deform the graph from (a) to  (e) because of the knot on the leg which cannot be undone. } \label{embno}
\endfig

More specifically, in the case of embedded framed graphs,  the 1$\rightarrow$3 Pachner move can always be
carried out in the embedded case without any further restrictions (i.e., if and only if it can be performed in the non-embedded case). The
2$\rightarrow$2 Pachner moves can be performed in the embedded case if it can be performed in the non-embedded case \emph{and} if the leg linking the two nodes is not knotted. The 3$\rightarrow$1 Pachner move can be performed in the
embedded case if it can be performed in the non-embedded case \emph{and} if the legs connecting the three nodes together are not knotted
\emph{and} there is no leg running through the hole formed by these three legs.

%We see therefore, that a subgraph such as the one presented in in
%\fg{embno} a) can never spread to the rest of the graph as our
%handles and cross-caps could in the non-embedded case. This is
%because, even though the two vertices

We therefore conclude that the knots such as the one shown in
\fg{embno} can never spread or propagate throughout the graph and always stays
localized on the graph.

%%%%%%%%%%%%%%%%%%%%%%%%%%%%%%%%%%%%%%%%%%%%%%%%%%%%%%%%%%%%%%%%%%%%
\section{Particles and their Propagation and Interaction in the 2d Case}

Having found the conserved quantities in 2d, in this Section we make some brief comments on the possible interpretation of these as particles or pseudo-particles.

The first problem we run into in trying to identify particles is actually defining what a particle is. Furthermore something that may look like a particle or quasi-particle at the microscopic level may not have the properties of a particle in an effective low-energy limit of the theory or model and vice versa.  This is why we chose to concentrate our efforts on conserved quantities in this paper as they will still be conserved in some effective limit.

At least with the Basic Rule, we only have a limited number of non-trivial structures and so we can formulate some sort of pseudo-particles with interactions and propagation. The only non-trivial structures with the Basic Rule are handles and cross-caps. We recall that the conserved quantity in the Basic Rule is the number of cross-caps plus two times the number of handles.

The very interesting properties of
these pseudo-particles is that, as absolutely any subgraph, they can
propagate, simply expanding the ``space" behind the particle by a
1$\rightarrow$3 Pachner move and contracting the space in front of
it by a 3$\rightarrow$1 Pachner move. This is very nice since simply
evolving the space through all possible Pachner moves will make the
particle sample all possible paths closely simulating a path
integral.

The other very nice property is that our two types of
pseudo-particle, the handles and the cross-caps can interact: if one end of a
handle crosses a cross-cap the handle splits into two cross-caps
(this does not change the global topology). We thus have some sort
of decay process of handles into two cross-caps and, obviously, also the
reverse process is possible.  This process is shown in
\fg{decay} .
\begin{figure}
\begin{center}
\scalebox{.8}{
\includegraphics{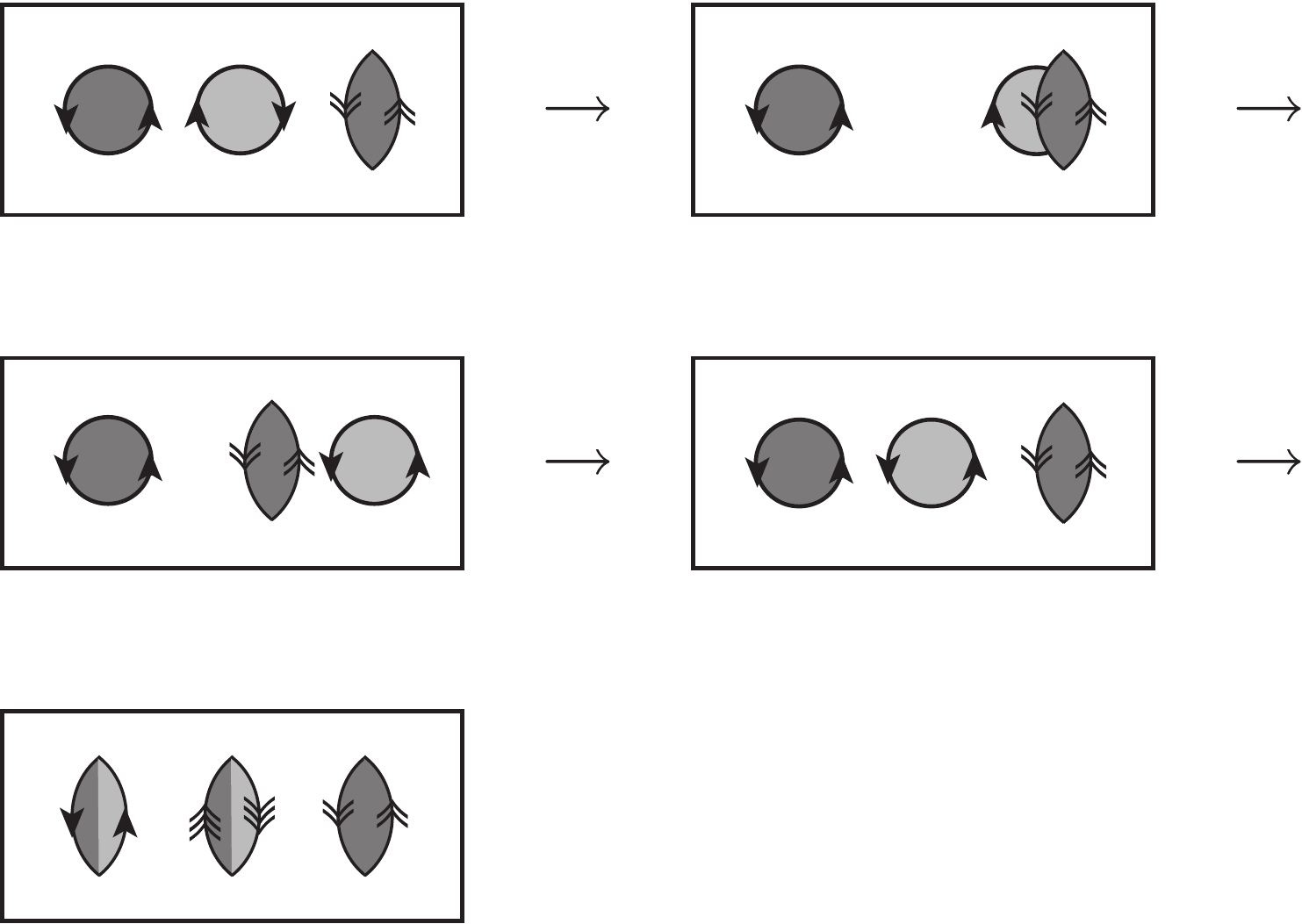}}
\end{center}
\capt{The topological process by which a handle ``decays" into two
cross-caps: one end of a handle goes through a cross-cap. The two
different greys represent two distinct holes in the surface.
} \label{decay}
\endfig

%%%%%%%%%%%%%%%%%%%%%%%%%%%%%%%%%%%%%%%%%%%COMPARE%%%%%%%%%%%%%%%%%%%%%%%%%%%%%%%%%%%%%%%%%%%%%
%-------------3 D--3 D--3 D--3 D--3 D--3 D--3 D--3 D--3 D--3 D--3 D--3 D--3 D--3 D--3 D--3 D--3 D--3 D--3 D--3 D--3 D--3 D--3 D--3 D--3 D-%

%%%Para on expectations of 3d case
\section{Expectations for the 3d Case}

%for Basic Rule
The theorem which applies in the Basic Rule for the 2d case equally applies to the 3d case.  Therefore, in the Basic Rule of the 3d Case, the only conserved quantity that will be conserved is the global topology. In three dimensions, the Prime Decomposition Theorem tells us that what will be replacing the handles and cross-caps of the 2d Case will be the trivial bundle $S^2 \times S^1$, the non-orientable $S^2\ $ bundle over $S^1$, and irreducible manifolds. This same theorem also suggests that the number of each irreducible prime manifold will be an exactly conserved quantity while the two different types of $S^2\ $ bundles over $S^1\ $ can transform into one another like the handles and cross-caps of two dimensions and only a certain combination of them is conserved analogously to the handles and cross-caps.

In addition, in 3d, the ``topological conical defects" mentioned in section \ref{SectionLine} will start to play a role and likely provide other possibilities for matter degrees of freedom. What is interesting in that case is that the points exhibiting ``topological conical defects" will form of string-like structures \mbox{\cite{DePietri:2000ii}} hinting at matter being fundamentally string-like.

%For Wan's rule
The results in two dimensions suggest that the conserved quantities for Smolin \& Wan's rule in three dimensions (other than the global topology) are
single tetrahedra with some of their faces and edges identified so as to have non-trivial topology.

Furthermore, it is worth noting that all of what has been done in \cite{SmoWan07} can be exactly transposed to the non-embedded Smolin \& Wan's Rule context since these papers did not, in actual fact, make use of the restrictions on Pachner moves that arise from an embedded setting.

%For Fotini's rule
As for the 3d case of the Closed Rule, we can already tell that something similar to the 2d case will occur. So that none of the tetrahedra composing the conserved structure can be expanded via the 3d \ad\ Pachner move, all these tetrahedra must have some identifications to make the topologically non-trivial. Hence we can expect that the conserved quantities of the 3d Closed Rule are a subset of the finite set of single tetrahedra with identifiactions and all possible multi-tetrahedral structures that can be built from the tetrahedra of the subset. We also expect the environmental dependance of some conserved quantities to persist in three dimensions.
%For the embedded rule

%--C O N C L U S I O N----C O N C L U S I O N----C O N C L U S I O N----C O N C L U S I O N--%
\section{Conclusions}

We studied three different models of evolving graphs and simplicial complexes  based on the Loop Quantum Gravity and Spin Foam framework: the Basic Rule, Smolin \& Wan's Rule and the Closed Rule.  Each of the three models can be considered either in the abstract case or the embedded case. In this paper we mostly focused on the abstract case where we found the conserved quantities for each model. The conserved quantities are subject to the hierarchy of the rules. That is to say that the the conserved quantities of the Basic Rule will also be conserved quantities of Smolin \& Wan's Rule and the conserved quantities of Smolin \& Wan's Rule will also be conserved quantities of the Closed Rule.

The quantitity which is conserved for all three Rules is the global topology which can be characterized by a single number, $T$:
\begin{equation}
 T= 2\times\mbox{number of handles} +\mbox{number of cross-caps}.
 \end{equation}
  In general we have the following conserved quantities:
\bite
\chew Basic Rule: $T$.
\chew Smolin \& Wan's Rule: $T + $ number of single-simplex cross-caps.
\chew Closed Rule: $T + $the number of single-simplex cross-caps $+$ number of each structure whose building blocks are triangles with at least two vertices identified.
\eat
It is interesting that the Closed Rule has the special feature that some of the conserved quantities are only conserved in a given type of environment. This seems very unusual but is not necessarily unheard of: in elecromagnetism the existence of a magnetic monopole implies the quantization of electric charge. It is such a dependance on the environment that is exhibited in the Closed Rule.

We also saw that,  in the Basic Rule,  handles and cross-caps can be seen as pseudo-particles that interact via the process of a handle and two cross-caps decaying into three cross-caps or vice-versa. In Smolin \& Wan's Rule, there is only one extra locally conserved quantity, so we may obtain a limited and finite particle spectrum.  The Closed Rule, in contrast has an infinite number of conserved quantities and so, naively, one would expect an infinite number of particles in such a theory. In a more realistic theory, the legs or lines of the graph will be labeled, and, depending on how the labelling is done and how it changes with evolution, the labeling can be the source of more conserved quantities, in particular these may lead to conical defects.

Finally, as mentioned in the introduction, geometrical conical defects, combined with the topological conserved quantities we found may correspond to topological geons which can exihibit a variety of different spin statistics, including the standard bosonic and fermionic statistics.\cite{Dowker:1996ei}  We have not attempted to relate the conserved quantities we found with other notions of 2d particles that can be found in the literature (see, e.g., \cite{FreKowSta08} and references therein).

%--A C K N O W L E D G E M E N T S----A C K N O W L E D G E M E N T S----A C K N O W L E D G E M E N T S--%
\section{Acknowledgements}

We would like to thank Martin Green and Yidun Wan for very useful suggestions discussions and are grateful to the Research Lab for Electronics at MIT where this work was completed for its hospitality.
One of the authors is supported by an NSERC-PGS scholarship and the project itself was partially supported by NSERC and a grant from the Foundational Questions Institute (fqxi.org).  Research at Perimeter Institute for Theoretical Physics is supported in part by the Government of Canada through NSERC and by the Province of Ontario through MRI.

%%%%%%%%%%%%%%%%%%%%  BIBLIO  %%%%%%%%%%%%%%%%%%%%%%%%%%%%%%%%

\newpage
%---------A N N E X E -------%
\section{Appendix A: Equivalence between adding a cross-cap and performing a surgery.}

The following surgery is equivalent to adding a cross-cap to the manifold:
\begin{enumerate}
\item
 Consider a trivial portion of a manifold with a disc removed from it ( the dark grey area in \fg{surgcross}(a)).
\item
Cut the portion of the manifold in half while remembering where we
will have to glue it back later.  (\fg{surgcross}(b): the simple grey arrows on the two
top edges mean that the edges must be identified so that the
arrows point in the same direction. The same applies for the double
arrows on the bottom edges. The two points marked with a grey circle must be
identified and similarly for the two points marked with a
grey star.)
\item
Place the cross-cap or M\"obius strip in the void between the two
parts of the manifold without gluing it yet and we continuously
deform the border of the disk (\fg{surgcross}(c)).
\item
Cut the cross-cap in the middle
while marking the points and edges which must be identified (\fg{surgcross}(d): Note
that the points marked a white circle and the points marked by a square are in fact the same points because they are the point
lying at the end of the single black arrow.)
\item
Glue half of the edge of the cross-cap with the
right half of the edge of the removed disk. (\fg{surgcross}(e):  This identifies the
points marked by a black star with the points marked by a grey star and the
points marked by a black circle with the points marked by a grey circle.)
\item
 Glue the other half of the edge. (\fg{surgcross}(f): The identification of the circle
with the star and the square with the triangle requires us to flip
the cross-cap around before making the gluing.)
\item
Continuously deform
the previously obtained manifold (\fg{surgcross}(g)).
\item
 Finally, glue back together
all the identified edges with the exception of the edge identified
with the triple black arrow (\fg{surgcross}(h)).
\end{enumerate}
We have thus achieved the prescription
of \fg{surgery} for adding a cross-cap to a 2-manifold.

\begin{figure}
\begin{center}
\begin{equation}
    \begin{array}{rcrc}
    a)&
    \mbox{\includegraphics{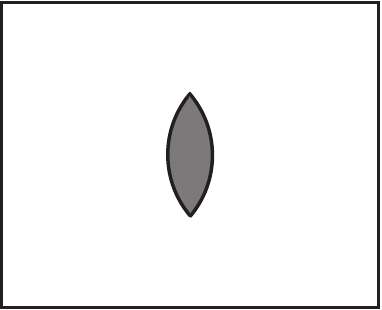}}&
    b)&
    \mbox{\includegraphics{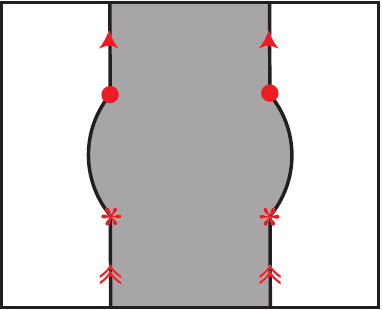}}\\
    c)&
    \mbox{\includegraphics{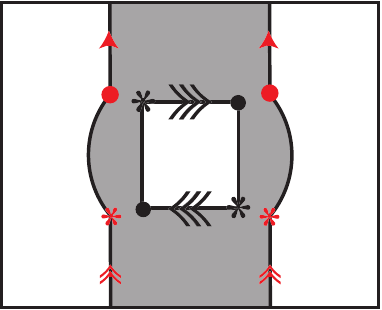}}&
    d)&
    \mbox{\includegraphics{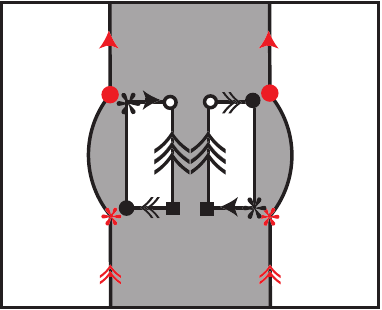}}\\
    e)&
    \mbox{\includegraphics{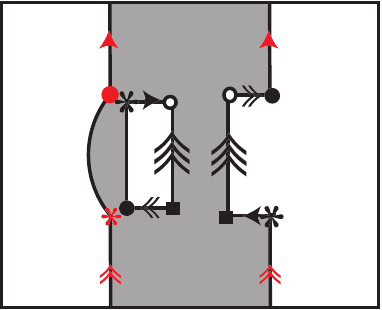}}&
    f)&
    \mbox{\includegraphics{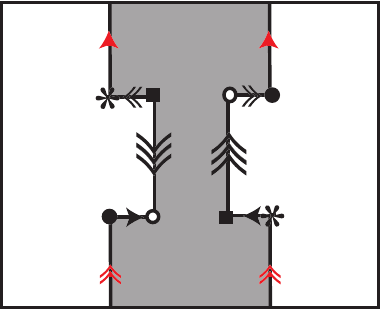}}\\
    g)&
    \mbox{\includegraphics{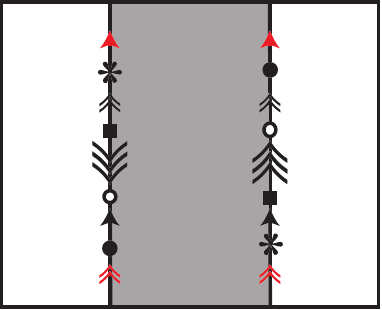}}&
    h)&
    \mbox{\includegraphics{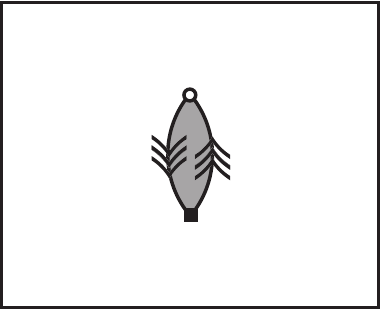}}
    \end{array}
      \nonumber
\end{equation}
\end{center}
\capt{This surgery is equivalent to adding a cross-cap.}
\end{figure}
\label{surgcross}

\section{Appendix B: Forbidden \bb\ Pachner move under Smolin \& Wan's Rule.}

\begfig
\begin{center}
\scalebox{.7}{
\includegraphics{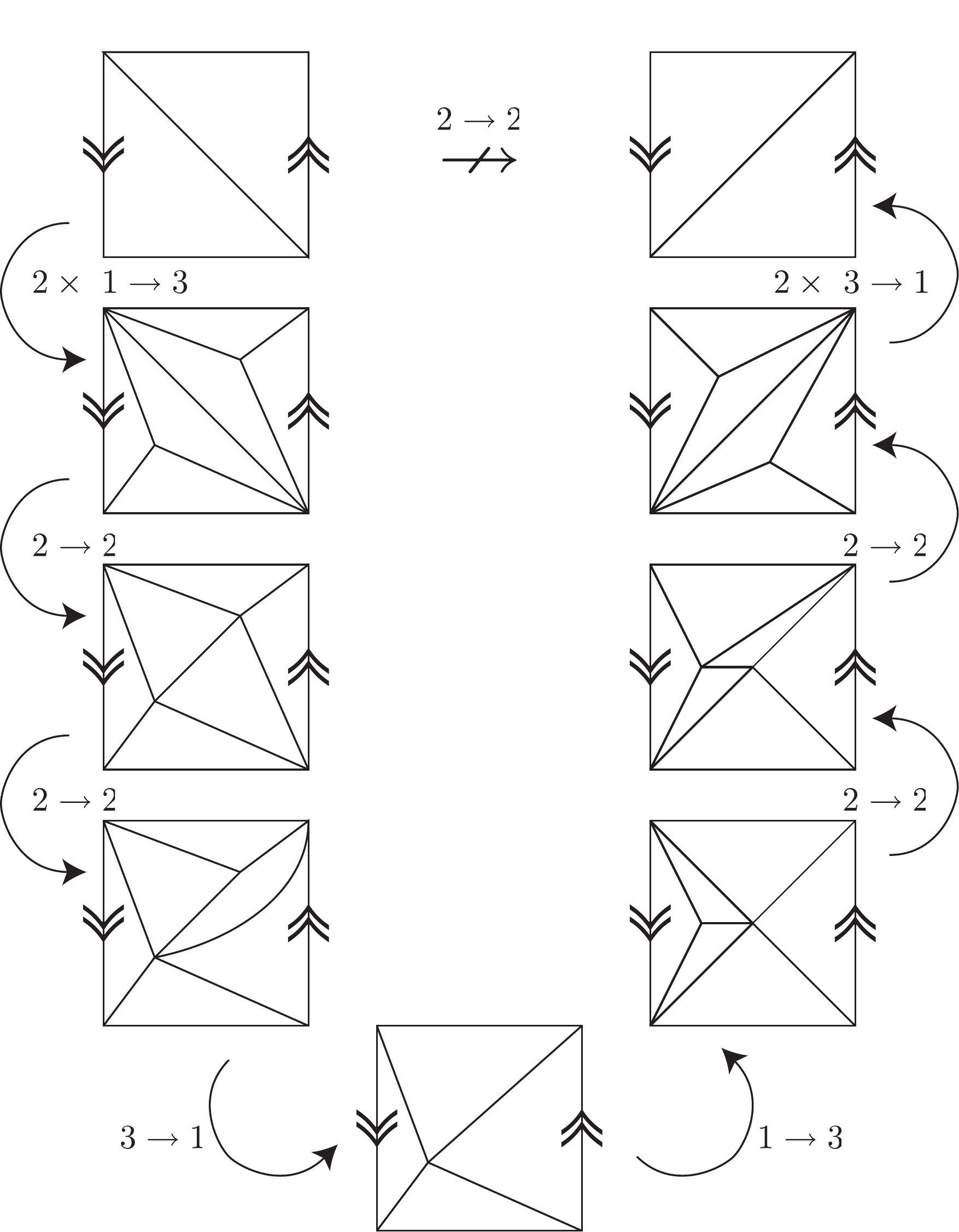}}
\end{center}
\capt{The forbidden \bb\ move.
The two edges identified with the double arrows are
identified such that the arrows match.} \label{wananex}
\endfig

The \bb\ Pachner move shown in \fg{wan22}(b) is forbidden under
Smolin \& Wan's Rule because the dual triangles form a M\"obius
strip whose interior is still a M\"obius strip and therefore not
homeomorphic to a disc. We can nevertheless achieve the same result
through a sequence of Pachner moves which are allowed by Smolin \&
Wan's Rule. Such a sequence is shown in \fg{wananex} in the triangulation
picture.

\section{Appendix C: The different ways to link link two simplices along two edges.}

Leg crossings in 2-node-subgraphs are equivalent to the same
subgraph without leg crossings but with extra line crossings.  The
possible crossing of the legs can be removed by
``rotating" one of the two central nodes along the horizontal axis.
This is shown in \fg{2dundobigperms}.
The two-line permutations at the extreme left and right of
\fg{2dpermini} can also similarly be removed  by
``rotating" the node which connects to our subgraph through these
permutations. This is shown in \fg{2dundosmallperms}.
This means that, for the local degrees of freedom, we can restrict
ourselves to consider only subgraphs of type shown in \fg{2dperm}.
This leaves us with only four possibilities as shown in
\fg{2dparts}, but, in fact, two of them are the same, (b) and (c). They
are simply different orientations, two different ways to ``embed"
the subgraph in the larger graph.

One can easily see that the same procedure also works in the
embedded case to remove any braidings of the legs at the cost of
extra line braidings. The fact that this works both in the
embedded and unembedded case is a particularity of the 2d case ($n=2$,
three-legged nodes). In the 3d case ($n=3$, four-legged-nodes),
 the leg crossings of the equivalent configuration can be
removed at the expense of extra line crossings in the unembedded
case.  However,  in the embedded case, a generic braiding of the legs is
not equivalent to an unbraided legs with extra line braidings.

The simplicial dual to the configuration of \fg{2dparts}(a)
is topologically only a 2-ball and therefore nothing
special as it does not change the topology of the graph within which
it is put. A rendition of the dual triangulation is given in
\fg{2dcone}(a) where the subgraph in \fg{2dparts}(a) is dual to a cone.

The configuration given in \fg{2dparts}(d) is more interesting, its
dual consists of two triangles joined together so as to form a
cylinder. This makes it into a micro-wormhole linking
together two regions of the whole graph. This is shown in \fg{2dcone}(b).

Finally, the configuration given in \fg{2dparts}(d) (and analogously
for \fg{2dparts}(c) ) has for dual the Moebius strip as is shown in \fg{moebius}.
This is also what we will call a cross-cap: cutting a disc out of a
2-manifold and gluing the border of the surface in \fg{moebius}
 to the border of the hole formed by
removing the disc adds a cross-cap. The last procedure is exactly
equivalent, topologically, as the procedure described in
\fg{surgery}. That this is the case is shown in \fg{surgcross}  of Appendix A.

\begin{figure}
\begin{center}
\scalebox{1}{
\includegraphics{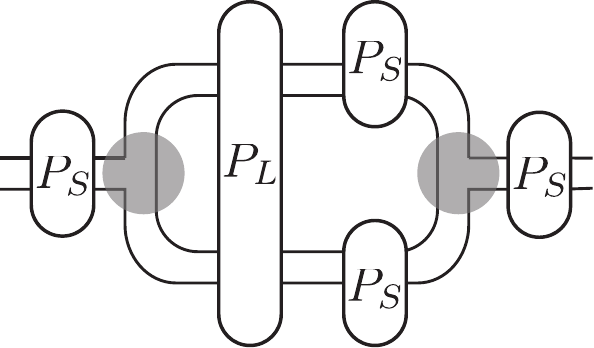}}
\end{center}
\capt{The most general line graph representation of how two simplices can be joined along two edges} \label{2dpermini}
\endfig

\begin{figure}
\begin{center}
\scalebox{.7}{
\includegraphics{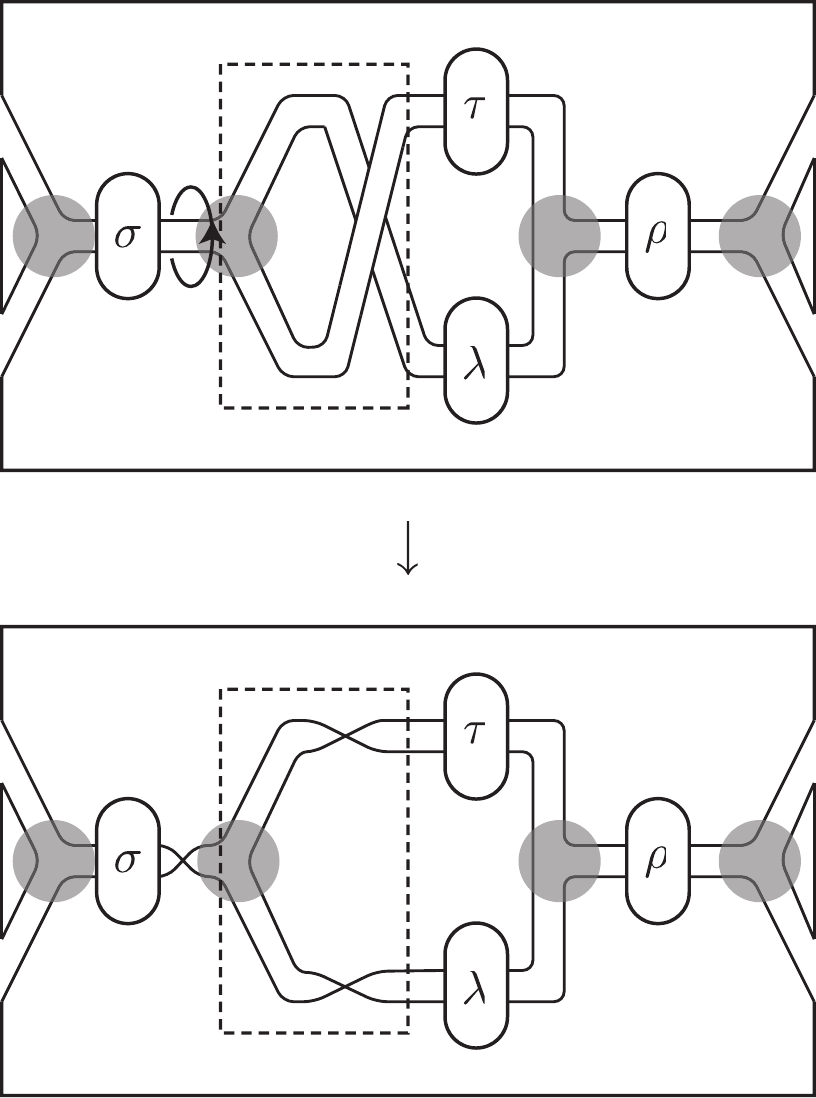}}
\end{center}
\capt{To undo the leg crossing, we can think of the graph as
embedded in $\R^3$ (even though it is not) and then rotate the
subgraph inside the dotted lines by $\pi$ along the leg coming out
of the dotted box on the left.} \label{2dundobigperms}
\endfig

\begin{figure}It seems possible, also,
\begin{center}
\scalebox{.7}{
\includegraphics{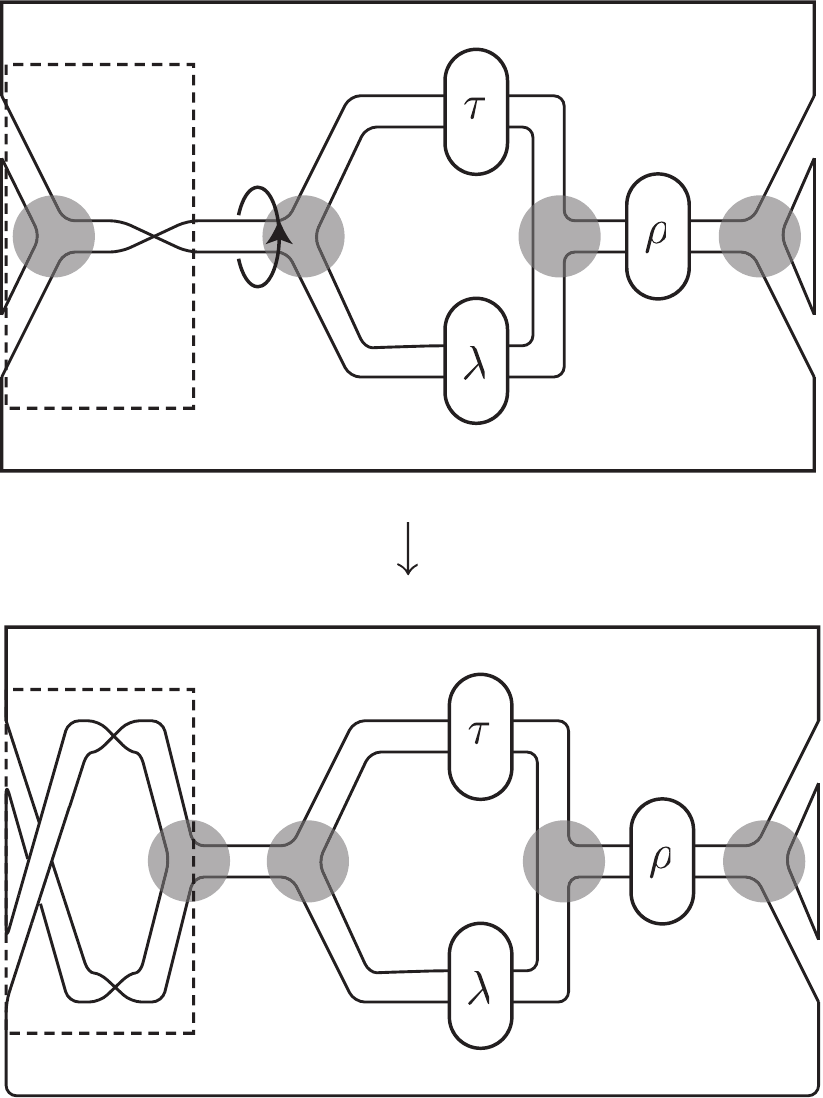}}
\end{center}
\capt{To undo the line crossing, we can think of the
graph as embedded in $\R^3$ (even though it is not) and then rotate
the subgraph inside the dotted lines by $\pi$ along the leg coming
out of the dotted box on the right. The same can be done to undo
$\rho$ if necessary. } \label{2dundosmallperms}
\endfig

\begin{figure}
\begin{center}
\scalebox{1}{
\includegraphics{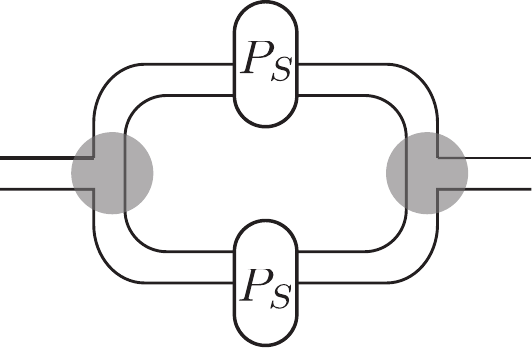}}
\end{center}
\capt{The local degrees of freedom.} \label{2dperm}
\endfig

\begin{figure}
\begin{center}It seems possible, also,
\begin{equation}
\begin{array}{cccc}
a)&
\scalebox{1}{
\includegraphics{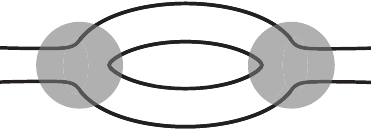}}&
b)&
\scalebox{1}{
\includegraphics{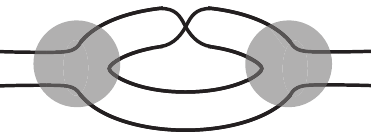}}\\
c)&
\scalebox{1}{
\includegraphics{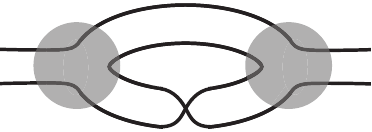}}&
d)&
\scalebox{1}{
\includegraphics{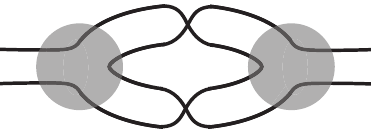}}
\end{array}
\nonumber
\end{equation}
\end{center}
\capt{The four possible configurations of the local degrees of
freedom of the pseudo-particles. Notice that (b) and (c) are locally the same,
 they only have a different orientation with respect to the
bigger graph. } \label{2dparts}
\endfig

\begin{figure}
\begin{center}
$a)$
\scalebox{.3}{
\includegraphics{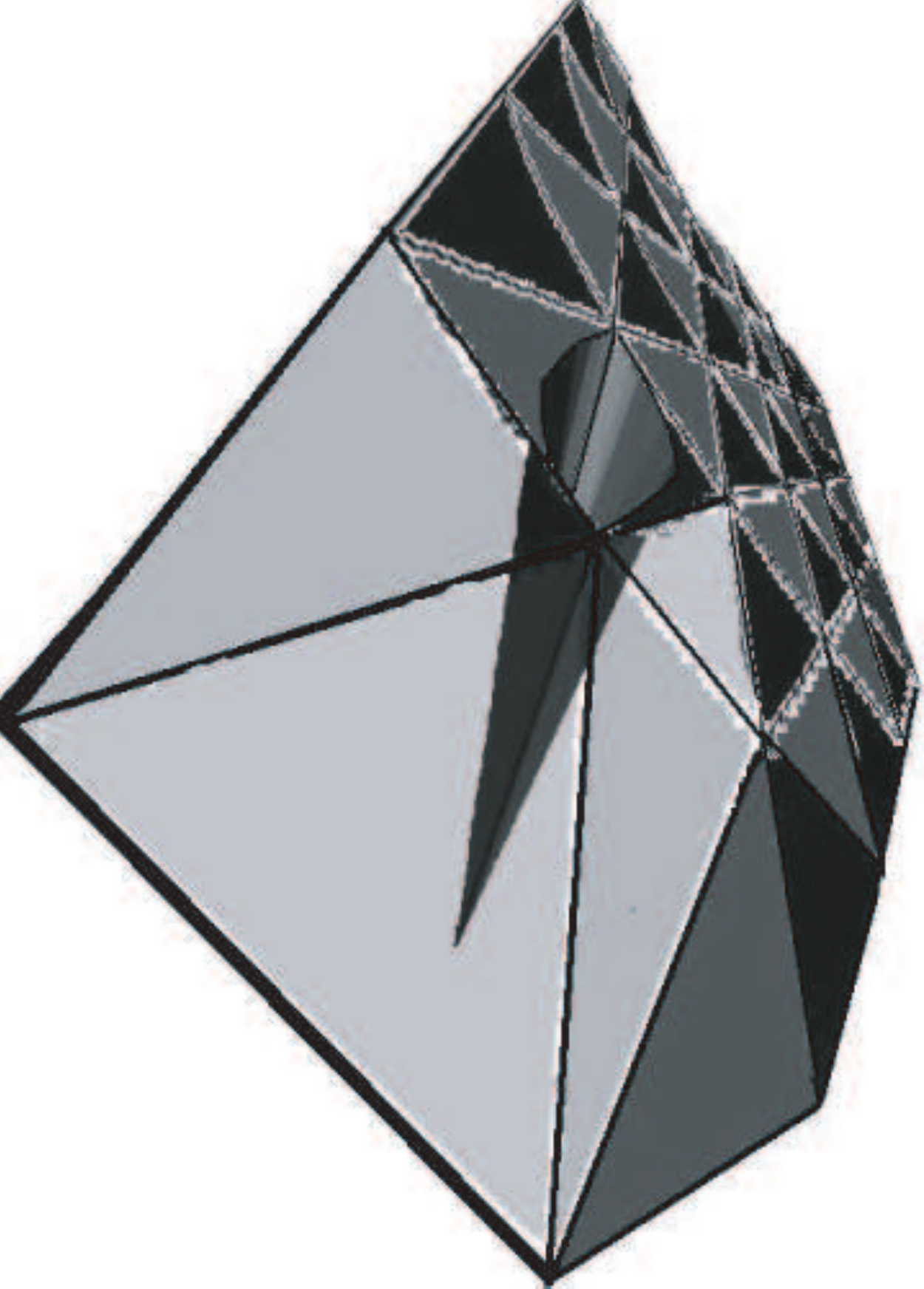}}
\qquad
$b)$
\scalebox{.3}{
\includegraphics{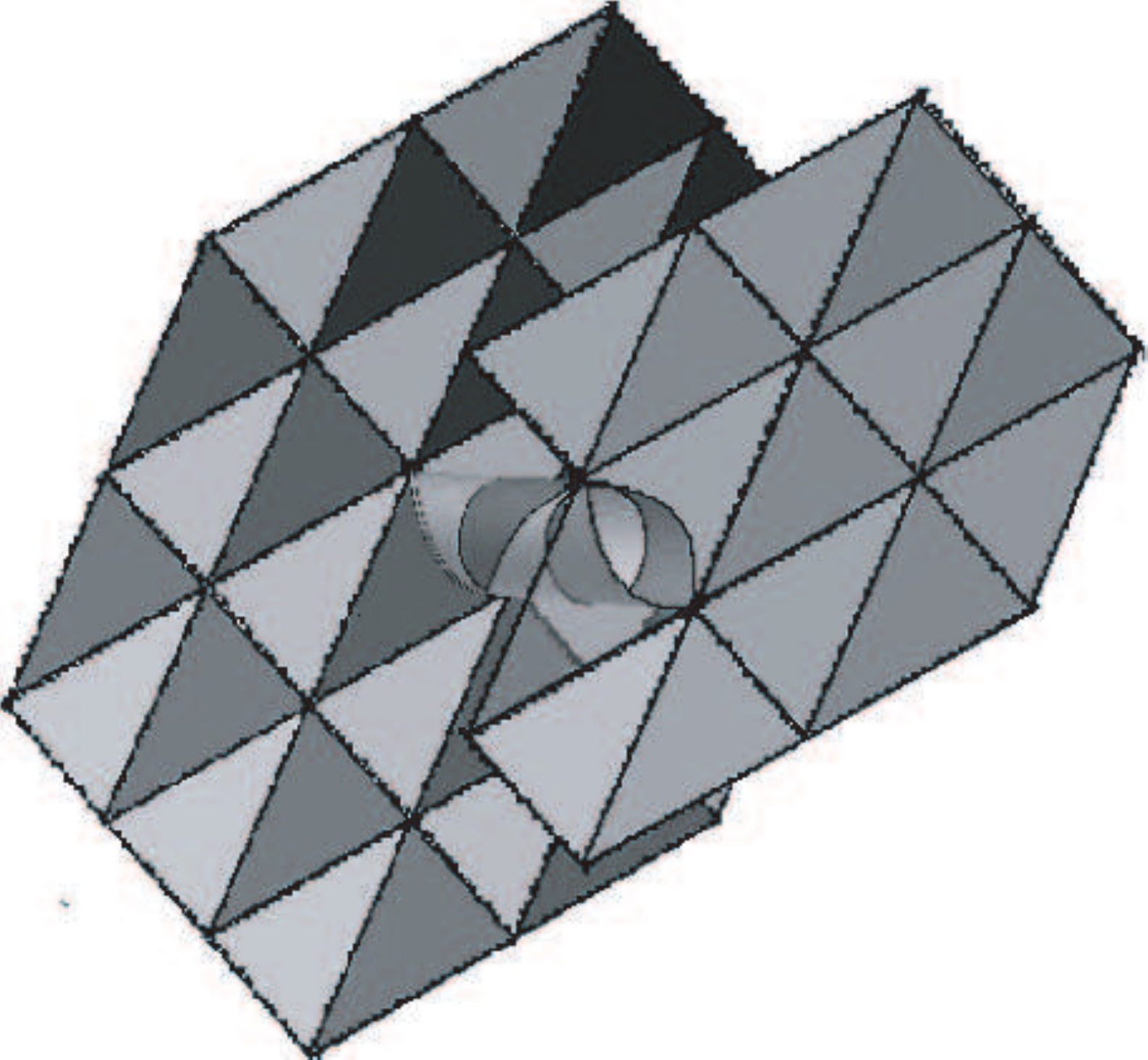}}
\end{center}
\capt{a)  The triangulation  dual  to the embedding of the
subgraph of \fg{2dparts}(a) in the graph dual to $\R^2$ shown in
\fg{dual2d}. The subgraph of \fg{2dparts}(a) is dual to the 2
triangles forming the cone.
b) Drawing of the dual triangulation to the embedding of the
subgraph of \fg{2dparts}(d) in the graph dual to $\R^2$ shown in
\fg{dual2d}. The subgraph of \fg{2dparts}(d) is dual to the two
white and orange triangles at the centre of the cylinder.} \label{2dcone}
\endfig

\begin{figure}
\begin{center}
\scalebox{.9}{
\includegraphics{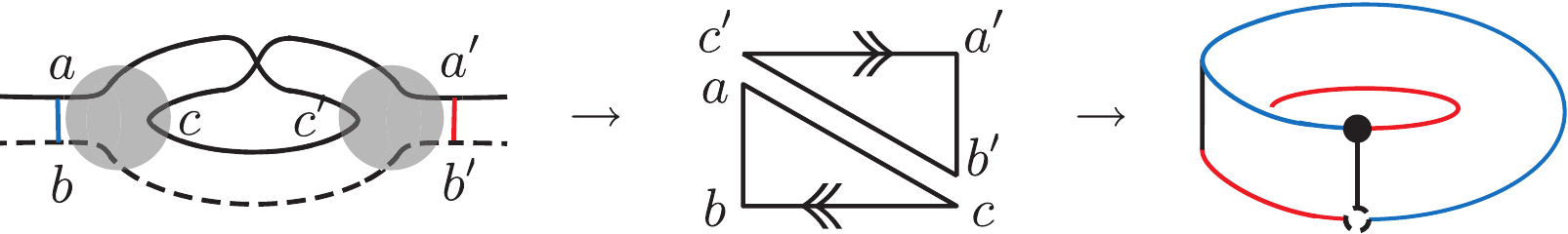}}
\end{center}
\capt{Drawing of the dual triangulation to the embedding of the
subgraph of \fg{2dparts}(b). The subgraph of \fg{2dparts}(b) is dual
to the M\"obius strip.} \label{moebius}
\endfig

\end{document}